\documentclass[11pt]{article}

\usepackage[margin=1in]{geometry}
\usepackage{amsmath,amssymb}
\usepackage{booktabs}
\usepackage{float}
\usepackage{graphicx}
\usepackage{microtype}
\usepackage{parskip}
\usepackage[T1]{fontenc}
\usepackage{lmodern}
\usepackage{natbib}
\usepackage{setspace}
\usepackage[colorlinks=true,citecolor=blue,linkcolor=blue,urlcolor=blue]{hyperref}

\newcommand{\E}{\mathbb{E}}
\newcommand{\Cov}{\operatorname{Cov}}
\newcommand{\Med}{\operatorname{Med}}

\title{Measuring Longer-Horizon Leveraged ETF Performance}
\author{Chris Murray\thanks{This note is a companion piece to Murray and Sammon, \emph{The Costs and Benefits of Leveraged ETFs} (SSRN: \href{https://papers.ssrn.com/sol3/papers.cfm?abstract_id=7133021}{abstract\_id=7133021}).}\thanks{Harvard Business School. Email: \href{mailto:cdmurray80@gmail.com}{cdmurray80@gmail.com}.} \and Marco Sammon\thanks{Harvard Business School. Email: \href{mailto:mcsammon@gmail.com}{mcsammon@gmail.com}.}}
\date{October 28, 2026}

\begin{document}

\singlespacing

\maketitle

\begin{abstract}
\begin{singlespace}

Many investors hold leveraged exchange-traded funds (LETFs) for more than one day. Because these products reset leverage daily, their longer-horizon returns need not equal the stated leverage multiple times the underlying return. We explain this gap using a return identity that separates what we call the convexity bonus, volatility drag, and product costs. We also discuss several common ways to measure LETF performance and highlight how these measures can produce misleading assessments of how real-world LETFs perform. We propose evaluating ex-post performance using realized terminal wealth, which measures how much invested wealth grew. For ex-ante analysis, we propose using expected log returns to quantify how fast capital compounds over time. We show that estimating expected cumulative return by compounding the historical arithmetic mean can rank investments backwards, report gains for funds that lose money, suggest that infinite leverage is optimal when expected returns exceed borrowing cost, and reward temporary price fluctuations that reduce LETF wealth. We argue that even if the distribution of daily returns is known by the econometrician, expected terminal wealth is sensitive to whether the model draws days independently or draws monthly or annual blocks of daily returns and can be driven by a small number of economically infeasible payouts.

\end{singlespace}
\end{abstract}

\newpage
\clearpage

\section{Introduction}

Leveraged exchange-traded funds (LETFs) have grown rapidly over the last several years, reaching over \$100 billion in assets under management. Given their rapid growth, it's increasingly important to understand how we should evaluate these products from an investor's perspective. This is complicated by the fact that, despite resetting leverage daily, many investors hold LETFs for more than one day. And, when held for more than one day, LETFs do not necessarily return their stated leverage multiple times the cumulative return of the underlying asset. For LETF investors, however, we believe the natural question is: how much did my investment grow?

The answer depends on the path of the underlying asset's returns, as well as fees, financing costs and the timing of investor flows. Whatever the drivers of returns, we believe that LETF investors ultimately care about the total change in their wealth from investing in these products. In this note, we clarify how to measure that change empirically and explain why log returns are more informative than other commonly used performance metrics, such as arithmetic average returns and expected terminal wealth.

Prior work has shown that the square of the underlying asset's cumulative return is a driver of LETF returns. In other words, LETF returns are a \emph{convex} function of the underlying asset's cumulative return \citep{tang2013solving}, holding volatility drag fixed. In this note, however, we highlight a decomposition of LETF performance that separates the effect of volatility drag from a force that always pushes in the opposite direction, which we call the \emph{convexity bonus}. The convexity bonus always adds to a long LETF's return (regardless of whether the underlying return is positive or negative), and is added to the payoff before volatility drag, fees, and financing costs take their cut of terminal wealth. We show that for a \(2\times\) strategy, the convexity bonus is the underlying asset's squared cumulative return, and for a \(3\times\) strategy it's three times the underlyer's squared return plus the cubed return. Importantly, this bonus can outweigh volatility drag when the underlying asset has an extremely large return, thus explaining how an \(L\times\) LETF can earn more than \(L\) times the underlying's cumulative return. As one example, in 2024, NVIDIA returned \(171\%\), while the \(2\times\) long NVIDIA LETF with ticker NVDX returned \(382\%\). In other words, the \(2\times\) LETF earned more than twice the underlying's return even net of fees, financing costs, and volatility drag.

Before going into the details, however, we would like to begin with a simple example in Table~\ref{tab:intro_variation} which shows just how misleading arithmetic average returns can be. The first row assumes that the underlying asset starts at \$100 and has a return of 0\% each period for two periods. Now, suppose a \(2\times\) LETF tracks the asset, resets daily, and has no fees or financing costs. That LETF will also have a cumulative return of zero given these returns. Now instead, suppose the asset starts at \$100, initially has a return of \(1\%\), and then a return of \(-0.9901\%\), returning to \$100, i.e., it again has a cumulative return of \(0\%\).

\begin{table}[htbp]
\centering
\small
\caption{Arithmetic Average Returns vs. Terminal Wealth Example}
\label{tab:intro_variation}
\begin{tabular}{cccc}
\toprule
Underlying price path & Underlying wealth & \(2\times\) wealth & Sample-mean benchmark\\
\midrule
\(100\to100\to100\) & \$100.0000 & \$100.0000 & \$100.0000 \\
\(100\to101\to100\) & \$100.0000 & \$99.9802 & \$100.0198 \\
\bottomrule
\end{tabular}

\vspace{0.4em}
\begin{minipage}{0.91\textwidth}
\footnotesize
\emph{Notes:} In this example, the \(2\times\) fund resets after each return and has no costs. The sample-mean benchmark column replaces the two realized fund returns with their arithmetic mean and compounds that mean over two periods.
\end{minipage}
\end{table}

The second row of Table~\ref{tab:intro_variation} shows that the \(2\times\) LETF earns \(2.0\%\) and then \(-1.9802\%\). Its arithmetic average daily return is positive, at \(0.0099\%\). Further, a hypothetical investment that earned this average return in both periods would grow from \$100 to \$100.0198. The actual investment in the LETF, however, falls from \$100 to \$99.9802. To summarize, this series of returns leaves the value of the underlying asset unchanged, reduces terminal wealth for an investor in the LETF, and increases the hypothetical wealth obtained by compounding the arithmetic average daily return. We refer to the hypothetical wealth obtained by compounding the arithmetic average daily return as the \emph{compounded sample-mean benchmark}.

We believe this two-period example makes the issue with arithmetic average returns obvious, but the misleading nature of arithmetic average returns does not depend on a reversal like the one observed in the second row of Table~\ref{tab:intro_variation}. Suppose instead that we wish to calculate another commonly used metric to evaluate LETF performance: \emph{expected terminal wealth}. To clarify, throughout this note, we use expected terminal wealth and expected cumulative return interchangeably because, for a fixed level of initial wealth, expected cumulative return equals expected terminal wealth divided by initial wealth, minus one. The compounded sample-mean benchmark estimates expected terminal wealth under independent, identically distributed (iid) returns. But, expected terminal wealth can also be computed by averaging terminal wealth across possible return histories generated by a statistical model of returns.

So, building on the example in Table~\ref{tab:intro_variation}, imagine we want to simulate expected terminal wealth for an investor in the asset or corresponding LETF. And suppose the underlying asset has two possible returns each period: \(1\%\) and \(-0.9901\%\). Further, assume these returns are equally likely independent draws each period. Independence is important here, as we are not baking reversals into the example like we were in Table~\ref{tab:intro_variation}. As the investment horizon increases, \textit{expected} (i.e., arithmetic average) terminal wealth grows for both the underlying asset and the LETF. Despite the fact that increasing the horizon increases expected terminal wealth, the underlying asset's long-run compound growth rate is zero. This rate describes how fast capital compounds along a sufficiently long realized path. Further, the LETF's long-run compound growth rate is negative and as the horizon increases LETF wealth converges (almost surely) to zero.

Maximizing expected terminal wealth can be rational in some circumstances, but we believe that long-horizon investors care about how their money compounds over many years, not just the average payoff across possible paths. This is a distinction that, despite its importance, many evaluations of LETFs overlook.

This is not the only issue with expected terminal wealth as a measure of LETF performance. Another issue is the fact that the expected return per period is not known to the econometrician. Instead, the model input of ``expected per period return'' is typically estimated using the historical arithmetic mean of a given realized return series, \(\bar r\). As we show, however, replacing the true population mean \(\mu\) with \(\bar r\) introduces another problem: the resulting compounded sample-mean benchmark is sensitive to the frequency at which returns are measured.

Further, even if \(\mu\) were known exactly, expected terminal wealth would measure the average terminal wealth across possible paths of returns, not the average long-run rate at which an investor's wealth compounds. Expected terminal wealth takes the arithmetic average of terminal wealth across possible paths and therefore can be heavily influenced by a small number of paths with economically infeasible terminal wealth. This is true even if returns are iid, because, by pure chance, a realized path may contain a long sequence of positive returns. Serial correlation changes the probability of such sequences and therefore the dispersion of terminal wealth. In fact, positive serial correlation will typically increase the dispersion of terminal wealth and can also increase expected terminal wealth without increasing the rate at which invested capital compounds on average over the long run. This happens because wealth on paths with a series of poor returns is bounded at zero, while paths with a series of positive draws can have extremely large terminal wealth.

Taking a step back, realized terminal wealth can be used to measure how much a fixed investment grew over a given period of time. It does not, however, account for the timing of investors' inflows and outflows, which ultimately determine investors' realized returns. As we discuss in \citet{MurraySammon2026}, the correct measure for that is investor profit and loss (P\&L), which accounts for when capital entered and left the fund. This is important because the average invested dollar need not earn the fund's full-period return which would be the case, e.g., if the fund attracted significant AUM after a series of very positive returns.

With all this in mind, our goal in writing this note is to clear up several common issues in evaluations of LETF performance. First, we show that compounding a fund's arithmetic average return -- i.e., applying the compounded sample-mean benchmark -- does not recover the realized growth of investor wealth. Second, even if the expected per period return were known exactly by the econometrician, it would not tell us the rate at which an investor's wealth is expected to compound over time in the corresponding LETF.

Third, we argue that a measure of long-horizon performance should be invariant to the frequency at which a given price history is sampled. Despite this, we show that estimates based on the sample arithmetic mean can differ depending on whether returns are measured daily, weekly, or annually, even though the realized terminal wealth is unchanged. A related problem arises in simulations where returns are drawn from a historical distribution: resampling returns one day at a time or in consecutive blocks (e.g., months or years) can produce different estimates of expected terminal wealth. The choice of block size changes which historical returns are grouped together, changing the dependence between returns and therefore expected terminal wealth. This result is counterintuitive because resampling at any frequency implies the same expected cumulative log growth over the investment horizon.

Fourth, the compounded sample-mean benchmark can rank investments backwards: a more volatile investment can produce less terminal wealth yet receive a higher ranking. Fifth, we show that a fund can lose money while its compounded sample-mean benchmark reports a large positive expected return. Sixth, under iid returns, if the arithmetic mean return exceeds financing costs, expected terminal wealth answers ``how much leverage?'' with ``always more.'' In fact, there is no finite leverage ratio that maximizes it. In many ways, this makes leverage like an avocado at your favorite lunch spot: adding it just makes everything better (although it will cost you). And in fact, expected terminal wealth continues to increase even when additional leverage reduces the investor's long-run compound growth rate.

Seventh, when compounded over long periods the expected wealth benchmark can be driven by low-probability economically impossible outcomes. For a \$10,000 investment in a 3X leveraged ETF on a volatile underlying asset compounding over 20 years, if we cap terminal wealth at roughly total global earthly wealth in 2026 we reduce expected terminal wealth by a factor of over 10,000,000.\footnote{Here the cap is \$600 trillion, giving capped
expected wealth of approximately \$64.8 million versus
\$683.6 trillion uncapped. Table~\ref{tab:gbm_tail_concentration}
instead uses a cap of \$1,000,010,000.}

Of course, there are sensible times to examine expected terminal wealth. However, we argue that expected terminal wealth is not a good benchmark for long-run compound growth. This is because we believe investors care about how money invested in a fund or LETF compounds over time, which is different than an investor who is trying to maximize the expected payoff from a one-off bet.

Tying all this together, we argue that working with log returns avoids all these aggregation issues with arithmetic average returns, both ex ante and ex post. A fund that lost money gets a negative average-log-return benchmark, i.e., there's no consolation prize for being volatile. Rankings on average log returns match the wealth actually produced. Changing the sampling frequency does not change the answer to ``which fund performed better?'' And expected cumulative log returns are the same whether the model draws days, weeks, or years from the same historical sample. When working with log returns, return variation still creates a penalty on compound growth, so the answer to ``how much leverage?'' is not automatically ``more.'' And a sextillion-dollar payoff in the distribution of possible returns no longer gets a sextillion-dollar vote. All of this is the result of compounding, not an assumption about investor utility.

Lastly, we discuss autocorrelation, which is often emphasized in discussions of LETF rebalancing and performance. Although autocorrelation matters for LETFs, we show that it also enters the expected terminal wealth of an investment in an ordinary stock, which does not rebalance. We argue, therefore, that it is important to separate the effects of rebalancing from the more general properties of compounded returns.

Of course, product design also matters for investor outcomes in LETFs. Leverage factors, reset frequency, financing costs, and fees all affect how the underlying asset's returns translate into investor wealth. This is particularly relevant as fund sponsors are now proposing LETFs that reset hourly rather than daily. And, hourly-reset funds provide a specific example of where the distinctions discussed above can matter for evaluating investor performance. If returns across reset intervals are independent and hourly variances add up to the daily variance, shortening the reset interval does not change the leading volatility-drag component of the underlying return needed for the LETF to break even with the underlying asset. Reset frequency can matter, however, if returns tend to continue or reverse across intervals.

Taking a step back, in \citet{MurraySammon2026}, our goal is to evaluate LETFs as investment products. And to do this, we explicitly measure what happens to the actual dollars people put into LETFs. To this end, we use realized log-return gaps, which reflect cumulative wealth differences, to compare how capital would compound in an LETF relative to its underlying asset. We use breakeven returns to measure the underlying performance needed to offset compounding, fees, and financing costs. Finally, our investor P\&L measure accounts for which products attracted capital and the timing of investor flows.

This note is related to several other papers on how to measure LETF performance. \citet{Bessembinder2026Volatility} studies the distribution of terminal wealth and shows that volatility need not reduce \textit{mean terminal wealth}, even as it widens the gap between mean and median outcomes. Holding leverage fixed and gross fund returns positive, a mean-preserving increase in return dispersion lowers the expected log growth of wealth. \citet{CarpenterLuWhitelaw2026} compare reset and non-reset strategies with the same initial leverage. Their continuously rebalanced strategy underperforms the non-reset benchmark in some states and outperforms it in others. In other words, they argue that neither strategy uniformly outperforms the other. Finally, \citet{lai2026lever} examine how financing costs, volatility, and serial dependence affect average terminal wealth and find that in a calibrated lifecycle portfolio-choice model there is substantial demand for LETFs. Our focus is instead on how realized returns, product design, costs, and investor flows determine the wealth earned by the capital actually invested in LETFs.

\section{The Basic Leveraged Return Identity}

We'll get started by looking at a self-financing strategy that rebalances continuously to keep \(2\times\) exposure to an asset that pays no dividends.\footnote{Equation~\eqref{eq:2xidentity} is exact for continuous rebalancing when \(S\) is a positive continuous semimartingale and \(\mathrm{RV}_{S,T}=[\log S]_T\) is the quadratic variation of its log price \citep{avellaneda2010path}. The full derivation of Equation~\eqref{eq:2xidentity} is in Appendix~\ref{app:continuous_identity}. For a daily-reset fund, replacing quadratic variation with a finite sum of squared returns gives a second-order approximation to Equation~\eqref{eq:2xidentity}. The exact formula for discrete-time rebalancing is in Appendix~\ref{app:discrete_identity}.} Let \(G_{L,T}\) denote terminal wealth per dollar initially invested in the strategy with leverage \(L\). Ignoring financing costs and management fees, terminal wealth per dollar invested is

\begin{equation}
G_{2,T}
=
\left(\frac{S_T}{S_0}\right)^2
\exp\left(-\mathrm{RV}_{S,T}\right)
=
\left(1+2R_S+R_S^2\right)
\exp\left(-\mathrm{RV}_{S,T}\right),
\label{eq:2xidentity}
\end{equation}

where \(R_S=S_T/S_0-1\) is the underlying asset's cumulative simple return and \(\mathrm{RV}_{S,T}\) is a realized-variance (RV) term measuring how much the asset's price fluctuates along its path between $S_0$ and $S_T$ \citep{avellaneda2010path}. \(\mathrm{RV}_{S,T}\) is commonly estimated by summing squared log returns. The factor \(\exp(-\mathrm{RV}_{S,T})\) captures what is commonly referred to as \emph{volatility drag}: higher realized variance multiplicatively reduces terminal wealth. This captures the idea that for a given cumulative return on the underlying asset, a more volatile path produces less terminal wealth for the LETF. Equation~\eqref{eq:2xidentity} does not depend on the distribution that returns are drawn from, nor does it require returns to be independent or normally distributed. Further, Equation~\eqref{eq:2xidentity} holds whether returns tend to continue, reverse, or are iid.

Although Equation~\eqref{eq:2xidentity} holds whether returns exhibit continuation, reversal, or neither, terminal wealth for a given \textit{realized} sequence of returns depends on the products of returns across reset periods. To see where these cross-products enter, consider a discrete-reset \(2\times\) strategy, with \(r_t\) denoting the underlying asset's simple return over reset interval \(t\). Expanding the discrete-reset wealth formula \(G_{2,T}=\prod_{t=1}^{T}(1+2r_t)\) and retaining terms up to quadratic order in the reset-period returns gives:
\begin{equation}
G_{2,T}
\approx
1+2R_S+2\sum_{i<j}r_i r_j.
\label{eq:2x_crossproducts}
\end{equation}

These cross-products do not represent a separate force, i.e., a force independent of the underlying asset's cumulative return and volatility drag. Rather, to second order, they equal what we call the \textit{convexity bonus} minus volatility drag. Writing Equation~\eqref{eq:2x_crossproducts} differently gives
\begin{equation}
2\sum_{i<j}r_i r_j
\approx
R_S^2-\sum_t r_t^2.
\label{eq:bonus_drag_crossproducts}
\end{equation}
Note that this rewriting has not removed the interactions between returns. It has bundled them into the convexity bonus associated with the underlying asset's cumulative return ($R_S^2$) and the offsetting volatility drag $\sum_t r_t^2$. In other words, Equation~\eqref{eq:bonus_drag_crossproducts} rewrites the cross-product expression in Equation~\eqref{eq:2x_crossproducts} in terms of the two forces we want to highlight throughout this note.

Taking logs of Equation~\eqref{eq:2xidentity} gives the same result in log returns. For leverage \(L\), we call the LETF's realized log return minus \(L\) times the underlying's realized log return the \emph{realized log-return gap}, \(D_T(L)\equiv\log G_{L,T}-L\log G_T\), where \(G_T=S_T/S_0\) is the underlying asset's gross return. The wealth identity and the log-return gap express the same thing in two different ways. Appendix~\ref{app:discrete_identity} gives the exact discrete-time identity for any general leverage factor $L$.

Someone new to LETFs might assume a 2X long LETF will have twice the cumulative return of its underlyer. Equation~\eqref{eq:2xidentity} shows the two forces that can make its actual return higher or lower than that. As mentioned above, we call the \(R_S^2\) term the \emph{convexity bonus}: it is never negative and depends only on the magnitude of the underlying asset's cumulative return. $\exp\left(-\mathrm{RV}_{S,T}\right)$ is typically referred to as \emph{volatility drag}, and it captures the penalty to terminal wealth from variation along the underlying asset's return path. Although the term ``volatility drag'' is somewhat imprecise (given that it measures quadratic variation rather than volatility itself), we believe it likely caught on because ``quadratic variation drag'' was too much of a mouthful. Importantly, the convexity bonus enters \textit{before} volatility drag is applied, so volatility drag reduces the entire payoff, including the convexity bonus.

Volatility drag gets most of the attention in prior literature and investor guidance on LETFs, while for some reason, the convexity bonus gets much less. As the name suggests, the larger the magnitude of the underlyer's cumulative return, the bigger the bonus. For a 2X strategy, a \(30\%\) move in either direction in the underlyer produces a bonus of \(9\) percentage points. A \(150\%\) gain in the underlying produces a bonus of 225 percentage points. This sounds unlikely to occur for an index LETF, but is feasible with a volatile enough individual stock.

The convexity bonus can outweigh volatility drag, as NVIDIA illustrates in Section~\ref{sec:nvda_convexity}. A simple example shows how. Suppose a stock gains \(10\%\) on two days in a row. Its cumulative return is \(21\%\), so simply doubling it gives \(42\%\). A costless daily-reset \(2\times\) fund earns \(44\%\). Using the exact discrete-time identity in Appendix~\ref{app:discrete_identity}, the convexity bonus contributes \(4.41\) percentage points and volatility drag removes \(2.41\). Positive autocorrelation can make two good or two bad days in a row more likely. But once those days have happened, the convexity bonus and volatility drag are what actually account for the LETF's return. Even in this simple example the volatility drag is still there. The convexity bonus just more than pays for it.

\section{Performance Measures}
\label{sec:measures}

As mentioned in the introduction, we believe the relevant performance metric for most investors is how invested capital compounds through time. Despite this, many different measures have been proposed to quantify LETF performance. To fix ideas, we start by defining the measures we discuss throughout this note.

Let \(r_t\) be the underlying asset's return over reset interval \(t\), so its gross return is \(1+r_t\). In other words, \(t\) indexes reset intervals: daily for a daily-reset fund, hourly for an hourly-reset fund, etc. Let \(L\) be the fund's stated leverage multiple. We'll use \(\bar r=\frac{1}{T}\sum_{t=1}^{T}r_t\) as the sample arithmetic mean of returns. For now we'll ignore fees and financing costs and assume that the underlying and fund gross returns are positive. This means the fund's return over each reset interval is \(r_{L,t}=Lr_t\), and its sample arithmetic mean is \(\bar r_L=L\bar r\). Finally, we'll define the gross wealth multiples of the LETF and underlying asset as
\begin{equation}
G_{L,T} \equiv \frac{W_{L,T}}{W_{L,0}}
=\prod_{t=1}^{T}(1+Lr_t),
\label{eq:letfwealth}
\end{equation}
and
\begin{equation}
G_T \equiv \frac{S_T}{S_0}
=\prod_{t=1}^{T}(1+r_t).
\label{eq:underwealth}
\end{equation}

Fixing initial wealth, the expected cumulative return equals expected terminal wealth divided by initial wealth, minus one. Table~\ref{tab:letf_performance_measures} defines several measures from the literature used to evaluate LETF performance. The middle column states the question we believe each measure answers. The last column gives examples of papers that use these particular measures.

\begin{table}[htbp]
\begin{center}
\footnotesize
\setlength{\tabcolsep}{3.5pt}
\begin{tabular}{p{0.19\textwidth}p{0.40\textwidth}p{0.32\textwidth}}
\toprule
Measure & Question & Examples in the literature \\
\midrule
$G_{L,T}$ & What did one dollar grow to over a given sample? & \citet{cheng2009dynamics}, \citet{Jarrow2010}, and \citet{charupat2011pricing} \\
$\E[G_{L,T}]$ & What is expected (arithmetic average) terminal wealth across possible return paths? & \citet{Bessembinder2026Volatility}, \citet{CarpenterLuWhitelaw2026}, and \citet{lai2026lever} \\
$\Med(G_{L,T})$ & What outcome is in the middle of the terminal wealth distribution? & \citet{Bessembinder2026Volatility} \\
$\E[\log G_{L,T}]$ & What is expected log wealth at a given horizon? & \citet{Bessembinder2026Volatility} \\
$\log G_{L,T}-L\log G_T$ & How far is the realized LETF log return from $L$ times the underlying log return? & \citet{avellaneda2010path} and \citet{MurraySammon2026} \\
Breakeven return & What must the underlying earn for the LETF to match it after costs? & \citet{MurraySammon2026} \\
Investor P\&L & How many dollars did investors actually gain or lose, given when capital entered and exited? & \citet{MurraySammon2026} \\
\bottomrule
\end{tabular}
\caption{LETF performance measures and the questions they address.}
\label{tab:letf_performance_measures}
\end{center}
\end{table}

The measures in Table~\ref{tab:letf_performance_measures}, other than investor profit and loss (P\&L), do not account for the timing of investor flows. As discussed in \citet{MurraySammon2026}, investor P\&L applies each return to the total capital actually invested by all investors at that time. Section~\ref{sec:applications} shows why this matters: a fund can earn a positive cumulative return even when its investors lose money because most of their money arrived after the fund's gains.

Throughout the note, we discuss the many issues with the compounded sample-mean benchmark. Before doing all that, however, we would like to highlight that there is a simple fix which avoids many of these issues. For a fixed initial investment, plugging the geometric mean return into the compounded sample-mean benchmark gives back the wealth actually earned. In this case, volatility ``drag'' hasn't disappeared: it's just getting smuggled in through the input, because the geometric mean is less than the arithmetic mean by roughly half the return variance. Table~\ref{tab:historical_input_benchmarks} compares benchmarks built from arithmetic means, geometric means, and average log returns. The last column divides historical wealth by benchmark wealth. A ratio of \(1\) means an exact match.

\begin{table}[htbp]
\centering
\small
\setlength{\tabcolsep}{5pt}
\begin{tabular}{@{}p{0.26\linewidth}p{0.23\linewidth}p{0.45\linewidth}@{}}
\toprule
Benchmark input & Implied benchmark wealth & Historical wealth / benchmark wealth \\
\midrule
\multicolumn{3}{@{}l}{\textit{Unlevered}} \\[4pt]
Arithmetic mean \(\bar r\) and serial covariance & \(\widehat{\mathrm E}[G_T]\) & \(\displaystyle \approx e^{-\frac{T}{2}[\sigma^2+(T-1)\bar\gamma_T]}\) \\[8pt]
Geometric mean \(\bar r_{\mathrm{geo}}\) & \(\widehat{\mathrm E}[G_T]^{*}\) & \(1\) \\[6pt]
Average log return \(\bar\ell\) & \(\exp\{\widehat{\mathrm E}[\log G_T]\}\) & \(1\) \\
\midrule
\multicolumn{3}{@{}l}{\textit{Levered}} \\[4pt]
Arithmetic mean \(L\bar r\) and serial covariance & \(\widehat{\mathrm E}[G_{L,T}]\) & \(\displaystyle \approx e^{-\frac{TL^2}{2}[\sigma^2+(T-1)\bar\gamma_T]}\) \\[8pt]
Geometric mean \(\bar r_{\mathrm{geo},L}\) & \(\widehat{\mathrm E}[G_{L,T}]^{*}\) & \(1\) \\[6pt]
Average log return \(\bar\ell_L\) & \(\exp\{\widehat{\mathrm E}[\log G_{L,T}]\}\) & \(1\) \\
\bottomrule
\end{tabular}
\caption{Benchmarks fitted to the same historical sample and horizon, with positive gross returns and costless leverage. The final column reports realized historical terminal wealth divided by the terminal wealth implied by each benchmark. Arithmetic rows use stationary fitted models and small-return approximations, neglecting higher cumulants. Here \(\sigma^2\) is the underlying return variance and \(\bar\gamma_T=\frac{2}{T(T-1)}\sum_{k=1}^{T-1}(T-k)\Cov(r_t,r_{t+k})\) is average covariance across distinct reset periods. The geometric mean is \(\bar r_{\mathrm{geo}}=G_T^{1/T}-1\), and the average log return is \(\bar\ell=T^{-1}\log G_T\), with analogous definitions using \(G_{L,T}\) for leveraged rows. Starred rows substitute geometric means, reproducing historical wealth rather than estimating expected wealth under the original fitted distribution. The volatility adjustment didn't disappear: it just moved into the input under a more respectable name: ``geometric mean.''}
\label{tab:historical_input_benchmarks}
\end{table}

Every performance comparison has two parts: what you're measuring and what you're measuring it against. Comparing a daily-reset LETF with a position levered once and never rebalanced isolates the effect of rebalancing. Comparing the LETF's log return with \(L\) times the underlying log return measures what we call the realized log-return gap, which isolates the effect of volatility on compounding. Comparing the LETF with the unlevered underlying asset asks whether the extra leverage paid off, but does not separately identify the effects of leverage and rebalancing. We bring this up because, as we show in the next section, the choice of what you're measuring anything against can matter just as much as what you're trying to measure in the first place.

\section{Comparing the Measures}
\label{sec:compare}

This section compares the main performance measures and explains why some of them can be misleading when describing past performance or forecasting future long-run compound growth, particularly for LETFs. We begin with the realized log-return gap and then examine how reset intervals can or cannot affect realized wealth. We next incorporate product costs through the breakeven return and distinguish expected terminal wealth from compound growth and from the return on a non-reset levered position.

\subsection{Volatility Drag and Leverage}

In the wealth identity, Equation~\eqref{eq:2xidentity}, we can see that realized volatility multiplicatively affects terminal wealth. For general leverage \(L\), this multiplier is
\begin{equation}
\exp\left[-\frac{L(L-1)}{2}\mathrm{RV}_{S,T}\right].
\label{eq:volatility_drag_multiplier}
\end{equation}
Taking logs to compute the log return gap turns it into a subtraction rather than a multiplicative factor, \(-\frac{L(L-1)}{2}\mathrm{RV}_{S,T}\). Either way, the drag scales with \(L(L-1)\), which is positive for leveraged long funds (\(L>1\)) and inverse funds (\(L<0\)).

Table~\ref{tab:volatility_drag_coefficients} shows that the volatility-drag coefficient in Equation~\eqref{eq:volatility_drag_multiplier} increases nonlinearly with leverage, and may be larger than expected for inverse funds. Specifically, a \(-1\times\) fund has the same second-order sensitivity to realized variation as a \(2\times\) long fund, while a \(-2\times\) fund has the same sensitivity as a \(3\times\) long fund.

\begin{table}[htbp]
\centering
\caption{Volatility-drag coefficient as a function of leverage factor.}
\label{tab:volatility_drag_coefficients}
\begin{tabular}{rrrrrrrr}
\toprule
\(L\) & \(0.5\) & \(1.25\) & \(1.5\) & \(2\) & \(3\) & \(-1\) & \(-2\) \\
\midrule
\(L(L-1)/2\) & \(-0.125\) & \(0.156\) & \(0.375\) & \(1\) & \(3\) & \(1\) & \(3\) \\
\bottomrule
\end{tabular}
\end{table}

\subsection{A portfolio with negative volatility drag}

Taking a closer look at Equation~\eqref{eq:volatility_drag_multiplier}, we can see that volatility drag is scaled by $L(L-1)$, which is positive for both levered long and inverse LETFs. But with $L$ between zero and one, this $L(L-1)$ term is negative, meaning volatility drag multiplicatively \textit{increases} terminal wealth relative to $(G_T)^L$ (this is why we were deliberately vague in saying ``multiplicatively affects'' above, as ``multiplicatively decreases'' wealth would be incorrect!). Initially, this seems really confusing, and one might wonder exactly what kinds of strange portfolios this applies to. This, however, is extremely common, e.g., a regularly rebalanced portfolio of stocks and cash.

As a specific example, imagine a daily-rebalanced 50/50 portfolio, i.e., a portfolio which is 50 percent stock and 50 percent cash or cash equivalents (e.g., T-bills). Also for now, let's assume the cash pays no interest. Plugging in $L=0.5$, the vol drag approximation becomes
\[
\exp\left[-\frac{L(L-1)}{2}\mathrm{RV}_{S,T}\right]
=
\exp\left[\frac{1}{8}\mathrm{RV}_{S,T}\right].
\]
Assuming a year with 252 trading days and daily volatility of $2.5\%$, and approximating the sum of squared returns by $252(0.025)^2$, volatility ``drag'' (now maybe it's clear just how misleading this language can be!) actually \textit{increases} terminal wealth by a factor of
\[
\exp\left[\frac{252(0.025)^2}{8}\right]
\approx 1.0199,
\]
or about $2\%$ relative to $(G_T)^{0.5}$. That's not life changing but remember this is over a given year. And, 2\% compounded over many years can be meaningful.

While the volatility ``drag'' \textit{increases} terminal wealth for our 50/50 portfolio, it's important to highlight that the convexity bonus now becomes a convexity penalty. For our 50/50 portfolio, the convexity bonus term becomes a penalty of
\[
1+\frac{R_S}{2}-\sqrt{1+R_S}.
\]
If the stock index gains $25\%$, simply taking half of its cumulative return gives $12.5\%$. However, the power-law component of the 0.5x strategy is $\sqrt{1.25}-1=11.8\%$. Thus, concavity creates a $0.70$ percentage-point shortfall relative to $12.5\%$, before accounting for the boost from volatility ``drag'', $\exp(\mathrm{RV}_{S,T}/8)$, which can offset some or all of that shortfall.

\subsection{Rebalancing, return patterns, and reset intervals}\label{freqsubsec}

Changing the reset interval changes which returns are grouped together. Importantly, even when daily returns are independent and identically distributed, a realized sample of returns can contain periods of continuation and reversal simply by chance. And importantly, the same returns can produce different terminal wealth depending on where the reset falls.

As one example, a monthly-reset strategy compounds levered monthly returns in the same way that a daily-reset strategy compounds levered daily returns. Within each reset block, leverage drifts away from its target, e.g., if returns are positive, the fund will drop below its leverage target until the next reset. The strategy restores the target leverage at the end of the reset block.

If a given LETF rebalances at any set of discrete intervals, the second-order volatility-drag term is the sum of squared returns over those intervals. Shorter reset intervals produce more returns to square in that sum, but each return is typically smaller. If returns are independent, these two effects approximately offset each other on average. If returns exhibit continuation or reversal, however, combining them over longer reset intervals introduces additional cross-products (as can be seen in Equation~\eqref{eq:bonus_drag_crossproducts}), so the effects need not offset. The next example illustrates this point.

Specifically, we construct an example to show that two LETFs tracking the same asset can earn different returns if they reset at different times. Suppose the underlying asset returns \(r_1\) and then \(r_2\). One fund resets after $r_1$ is realized, i.e., after the first period, the second LETF waits to reset leverage until the end of the second period. The difference in these LETFs' ending wealth is
\begin{equation}\label{eq:rearr}
\begin{aligned}
&(1+Lr_1)(1+Lr_2)
-\left\{1+L[(1+r_1)(1+r_2)-1]\right\}\\
&\qquad=L(L-1)r_1r_2.
\end{aligned}
\end{equation}
This shows that for long leveraged funds and inverse funds, resetting between $r_1$ and $r_2$ helps when both returns are in the same direction. Equation~\eqref{eq:rearr} also shows, however, that resetting between period 1 and 2 hurts when the second move is a reversal relative to the first. We believe this example also helps make the broader point that reset frequency can change volatility drag and the breakeven return without changing the path of the underlying stock. This happens because the leading term of the volatility drag approximation depends on the sum of squared returns \(\sum_t r_t^2\) on a given path, where $r_t$ is calculated over each reset interval.

\subsection{Breakeven returns and product costs}

Intuitively, we believe investors evaluating LETFs want to know whether the additional exposure provided by leverage overcame volatility drag and costs. In \citet{MurraySammon2026}, we summarize this comparison using what we call a breakeven return. Let \(G_{L,T,\mathrm{net}}\) denote terminal wealth after financing costs and management fees. We summarize the combined effect of these costs by \(c_T\), where \(e^{-c_T}\) is the ratio of net terminal wealth to the wealth the strategy would have produced in the absence of those costs.

Retaining the leading quadratic term from Equation~\eqref{eq:volatility_drag_multiplier} and adding costs gives
\begin{equation}
G_{L,T,\mathrm{net}}
\approx
(G_T)^L
\exp\left[
-\frac{L(L-1)}{2}\sum_{t=1}^{T}r_t^2-c_T
\right].
\label{eq:costloggap}
\end{equation}
The term \((G_T)^L\) contains the convexity bonus generated by leveraged exposure, while the exponential term captures the offsetting effects of volatility drag and costs. For a long fund with \(L>1\), the underlying cumulative simple return at which these forces balance and the LETF matches the unlevered asset is approximately
\begin{equation}
R_T^{*}
\approx
\exp\left[
\frac{L}{2}\sum_{t=1}^{T}r_t^2
+\frac{c_T}{L-1}
\right]-1.
\label{eq:breakeven}
\end{equation}

Specifically, Equation~\eqref{eq:breakeven} identifies the cumulative return at which the benefit of leveraged exposure, including the convexity bonus, offsets volatility drag and costs under the quadratic approximation. As expected, this breakeven return rises with the sum of squared reset-period returns, which drives volatility drag, as well as with fees and financing costs. Fees and financing costs multiplicatively shrink terminal wealth. For example, a \(2\times\) fund with a \(1\%\) annual fee and a \(4.5\%\) financing rate has total annual costs of approximately \(5.5\%\) of AUM. Over ten years, it retains \(e^{-10(0.055)}=57.7\%\) of hypothetical no-cost terminal wealth. This does not mean that \(\$100\) falls to \(\$57.70\): whatever wealth the no-cost strategy would have produced is multiplied by \(57.7\%\). Appendix~\ref{app:additional_applications}, particularly Appendix~\ref{app:long_run_costs}, reports historical cost decompositions and longer-horizon financing simulations. Appendix~\ref{app:moderate_leverage} considers a moderate-leverage, low-cost design.

\subsection{Expected terminal wealth and compound growth}

This section distinguishes the average payoff across possible paths from the rate at which capital compounds through time.

Under iid daily returns with \(\E[r_t]=\mu\),
\begin{equation}
\E[G_{L,T}]=(1+L\mu)^T.
\label{eq:expectedterminal}
\end{equation}
Holding the arithmetic mean fixed, volatility does not appear in expected terminal wealth. This result is central to \citet{Bessembinder2026Volatility} and \citet{CarpenterLuWhitelaw2026}: daily-reset leverage does not automatically reduce the \textit{arithmetic} average of terminal wealth across possible paths.

Compound growth is calculated by multiplying returns across periods rather than averaging them. Because log returns add, expected cumulative log return over \(T\) reset intervals is \(\E[\log G_{L,T}]=T\E[\log(1+Lr_t)]\). For small reset-period returns,
\begin{equation}
\E[\log(1+Lr_t)]
\approx L\mu-\frac{L^2}{2}\E[r_t^2].
\label{eq:expectedlog}
\end{equation}
Return variation ($\E[r_t^2]$) therefore enters the formula for compound growth in Equation~\eqref{eq:expectedlog}. This follows from taking logs of Equation~\eqref{eq:letfwealth}, not an assumption about investor preferences. More generally, compounding determines how one-period return distributions translate into long-horizon wealth distributions \citep{FaragoHjalmarsson2023}.

Separately, \citet{CarpenterLuWhitelaw2026} show that a continuously rebalanced leveraged strategy is self-financing and is not dominated by a non-reset leveraged position: it can outperform after large moves in either direction and underperform after more moderate moves. That result compares reset and non-reset strategies, but does not resolve the question of whether expected terminal wealth measures compound growth (which is what we argue investors ultimately care about).

The non-reset benchmark is a ``lever once'' strategy. And ``once'' really means \textit{once}, i.e., not once a year, not once a decade, but once, ever. Suppose you put down \$10,000 in 1970, borrowed another \$10,000, and bought \$20,000 of S\&P 500 exposure. The non-reset benchmark never borrows more to restore leverage. With dividends reinvested and financing costs ignored, by October 9, 2026, you would hold approximately \$8.23 million against the original \$10,000 debt. In other words, your terminal leverage (assuming you never repaid your original \$10,000) would be just \(1.0012\times\). This is what the non-reset benchmark means over that holding period.

Repeating a one-year ``lever-once'' calculation with \(2\times\) starting leverage each year silently introduces annual resetting. This is not ``lever-once'', whether performance is measured using expected wealth or expected log wealth. And once we're rebalancing, then all the math in Appendix~\ref{app:discrete_identity} applies, with the convexity bonus and volatility drag calculated using annual reset intervals.

\section{Expected Wealth as a Performance Benchmark}
\label{sec:expectedwealthbenchmark}

An ``ex ante'' benchmark still needs inputs. If the inputs come from historical returns, a natural question is: does the benchmark match the wealth those returns actually produced? If they don't, where do they come from, and how do they relate to the security's actual performance? Further, once we're drawing returns from the historical distribution, there are many other questions that need answering, e.g., should the model draw days independently, or draw whole weeks, months, years, or decades of returns? And finally, once we decide we're interested in an ex ante measure of expected performance rather than an ex post benchmark, should we use arithmetic expected returns or expected log returns (especially given that both can be calculated from the same model)? In our experience, many of the misleading results produced by some commonly used benchmarks come from overlooking the compounding penalty from return variation, and LETFs magnify that omission by design: a costless \(2\times\) fund doubles daily returns but quadruples their variance.

To fix terminology in this section, under iid returns calibrated to the historical arithmetic mean,
\begin{equation}
\widehat{\E}[G_{L,T}]=(1+L\bar r)^T.
\label{eq:historicalexpectedwealth}
\end{equation}

The subsections below examine how this benchmark compares with realized wealth, how extreme outcomes and leverage affect expected wealth, how temporary price movements and return-modeling choices change the comparison, and how trading costs enter the observed return path.

\subsection{Historical wealth, fund rankings, and losing investments}

Compounding the historical arithmetic mean to estimate \(\E[W_{L,T}]\) makes the investment look better than it actually was whenever returns vary. This isn't unique to LETFs, but leverage makes this mistake worse: expected returns scale with leverage, while variance scales with its square.

With positive reset-period gross returns,
\begin{equation}
G_{L,T}=\prod_{t=1}^{T}(1+Lr_t)
\leq (1+L\bar r)^T,
\label{eq:amgm}
\end{equation}
with strict inequality whenever the fund's returns are not literally the same every day. Every varying realized path therefore falls short of its own arithmetic-mean benchmark by construction, not because of bad luck.

The compounded sample-mean benchmark ranks funds by their arithmetic average returns, not the wealth they actually produced. A volatile fund with a lower geometric return can outrank a less volatile fund with a higher geometric return. In other words, this benchmark can make a loser look like a winner, and even rank losers above the winners.

A fund can lose money while its compounded sample-mean benchmark is positive. Knowing the true distribution of returns doesn't make this problem go away: a positive arithmetic expectation can sit alongside a negative long-run compound growth rate. With iid returns, expected wealth can keep rising while wealth along almost every realized path heads toward zero.

For wealth following geometric Brownian motion (GBM), with drift \(a\) and volatility \(\sigma_W\), expected wealth grows at rate \(a\), but the long-run log growth rate is \(a-\sigma_W^2/2\), so when the drift is positive but the variance penalty is larger, the expectation gets richer while the investment gets poorer. This $-\sigma_W^2/2$ term is the It\^{o} correction, and has nothing to do with investors (possibly) having mean-variance preferences.

Imagine a hedge fund manager who uses the compounded sample-mean benchmark when writing their annual investor letter. After losing \(10\%\) in a given year with daily log-return volatility of \(3.5\%\), the manager gets to report a positive estimated expected-return benchmark of approximately \(5.0\%\). Then next year the fund loses another \(10\%\), but daily volatility rises to \(4.5\%\), producing a benchmark return of approximately \(16.2\%\). The second letter is a victory lap: ``expected returns'' have improved substantially! Investors' \$100 has become \$81. The fund didn't get better at making money, the benchmark just rewarded the extra volatility.
\footnote{These illustrative estimates use the second-order
approximation \(G\exp(252\sigma_d^2/2)-1\), where \(G=0.9\)
is realized annual gross wealth and \(\sigma_d\) is daily
log-return volatility. The approximation neglects higher-order
terms and does not require fund returns to follow GBM.}

\subsection{A mean based on impossible outcomes}

Table~\ref{tab:gbm_tail_concentration} shows long-term results for wealth growth for a $\$10,000$ initial investment in a 2X and 3X LETF on a volatile underlying asset with a 20 year time horizon. We include both mean wealth \(\E[W_{L,T}]\) and capped mean wealth \(\E[\min(W_{L,T},\mathrm{cap})]\) where the cap is $\$1,000,010,000$, a $10,000,000\%$ return. The mean wealth, at least for the 2X, might at first appear plausible. But the columns on the right of the table show that this average is driven by the roughly $0.02\%$ of paths above the uncapped mean whose terminal wealth averages $\$866$ trillion, representing slightly more than common estimates of total wealth on earth in 2026. We claim (without proof, of course) that this is a strictly impossible economic outcome despite the fact that it is a mechanically true result of the mathematical model. The 3X manages to dial up ``already impossible'' by a factor of about 15 million: paths with terminal wealth averaging $\$13.5$ sextillion drive the mean. Meanwhile, arithmetic mean wealth implicitly assumes we are able to colonize the entire solar system over the next 20 years. Specifically, the geometric mean of terminal wealth (which equals median terminal wealth under this model) is only \(\$1.90\) for the \(2\times\) strategy and effectively zero for the \(3\times\) strategy. In other words, a typical \(\$10{,}000\) investment is almost entirely wiped out, even though a tiny number of astronomically large outcomes produce an enormous arithmetic mean.

Of course, positive skewness is a real feature of stock-market returns: most individual stocks underperform Treasury bills, and the relatively few winners account for most of the market's net wealth creation \citep{Bessembinder2018,BessembinderChenChoiWei2023}. Further, compounding itself can generate positive skewness, with lognormal wealth under GBM providing a familiar example \citep{FaragoHjalmarsson2023}. But a small number of actual winners driving observed market gains are different from economically impossible payouts driving a model's expectation. Apple becoming a multi trillion dollar behemoth is a real example of positive skew, but a \$10,000 investment paying \$13.5 sextillion in 20 years is (to us) implausible. Both distributions may be positively skewed, but evidence for the former does not make the latter economically credible.\footnote{Out of morbid curiosity we ran this calculation for the 4X LETF as well. This gives mean wealth \$2.8 quintillion, mean capped wealth of \$21.41, and geometric mean wealth of $\$4.67 \times 10^{-26}$. Said a different way, $99.9999999999373\%$ of paths are below the mean, but the remaining extreme-tail paths above the mean average $\$4.46\times10^{30}$ terminal wealth. Using \href{https://www.nasa.gov/missions/kepler/about-half-of-sun-like-stars-could-host-rocky-potentially-habitable-planets/}{NASA's conservative estimate} for potentially habitable planets in the Milky Way, we believe this could purchase roughly 25 million galaxies of global wealth, assuming each habitable planet has 2026 Earth wealth.}

To compare with a real-life example of an extraordinary winner, we backtested hypothetical daily-reset strategies using NVIDIA's (Ticker: NVDA) realized returns from January 3, 2006 through September 30, 2025, just shy of 20 years. An initial \$10,000 grew to \$6.68 million in NVDA, \$32.99 million in a costless \(2\times\) strategy, \$12.31 million in a \(2\times\) strategy with annual total costs of \(5\%\) of AUM, including expenses and financing, and \$454,000 in a costless \(3\times\) strategy. In reality, the 3X would almost certainly have faced an intraday swap close out on July 3, 2008, when NVDA had a close-to-close return of -30.7\%. NVDA was an extraordinary actual winner, but even the costless \(2\times\) strategy falls short of the average wealth on the paths driving the theoretical \(2\times\) mean by roughly 26 million times.

\begin{table}[htbp]
\centering
\scriptsize
\setlength{\tabcolsep}{3pt}
\resizebox{\textwidth}{!}{\begin{tabular}{@{}crrrrrrr@{}}
\toprule
& & & &
\multicolumn{2}{c}{Below the uncapped mean} &
\multicolumn{2}{c}{Above the uncapped mean} \\
\cmidrule(lr){5-6}\cmidrule(lr){7-8}
\(L\)
& \shortstack{Mean\\wealth}
& \shortstack{Mean wealth\\with cap}
& \shortstack{Geometric-mean wealth\\
\(W_0\exp\{\E[\log G_{L,T}]\}\)}
& \shortstack{Percentage\\of paths}
& \shortstack{Average\\wealth}
& \shortstack{Percentage\\of paths}
& \shortstack{Average\\wealth} \\
\midrule
\(2\times\)
& \$167.18 billion
& \$3.97 million
& \$1.90
& \(99.980713\%\)
& \$32.25 million
& \(0.019287\%\)
& \$866.62 trillion \\
\(3\times\)
& \$683.56 trillion
& \$39,672
& \(\$1.62\times10^{-10}\)
& \(99.99999494\%\)
& \$34.60 million
& \(0.00000506\%\)
& \$13.50 sextillion \\
\bottomrule
\end{tabular}}
\caption{Expected wealth and tail concentration for a \$10,000 investment over 20 years.}
\label{tab:gbm_tail_concentration}

\vspace{0.4em}
\begin{minipage}{0.95\textwidth}
\footnotesize
\emph{Notes:} The underlying follows GBM with expected daily log return \(0.0004\) and daily log-return volatility \(5\%\). Strategies maintain continuous constant leverage, with 252 trading days per year and no fees or financing costs. Geometric-mean wealth equals median wealth under this model. The capped calculation replaces terminal returns above \(10{,}000{,}000\%\) with that return, allowing ending wealth up to \$1,000,010,000, and no paths are discarded. All other columns describe the uncapped distribution. One sextillion is \(10^{21}\).
\end{minipage}
\end{table}

\subsection{Expected wealth wants unlimited leverage}

Under iid returns with arithmetic mean \(\mu>0\),
\begin{equation}
\E[W_{L,T}]=W_{L,0}(1+L\mu)^T.
\label{eq:benchmarkleverage}
\end{equation}
By inspection, we can see that more leverage makes this number bigger. If you impose a leverage cap, we can still see that being at that cap is optimal. And of course, if we remove that cap, there is no finite optimum level of leverage.\footnote{This is an unconstrained, costless benchmark. In discrete time, requiring \(1+Lr_t>0\) in every possible state can impose a finite leverage limit, and margin, termination, and borrowing constraints can bind earlier. With a constant financing rate \(r_f\), the per-period mean gross return becomes \(1+L\mu-(L-1)r_f\), so the monotonicity claim requires \(\mu>r_f\). The continuous-time GBM model also has increasing expected wealth without a finite leverage cap when the drift exceeds the financing rate.} This remains true whether \(\mu\) is known exactly or estimated from historical data.

Neither realized compound wealth nor expected log wealth have this flaw of automatically recommending infinite leverage. For realized compound wealth, the wealth identity in Equation~\eqref{eq:2xidentity} shows the competing forces: greater exposure can add to the payoff, but volatility drag takes more and more back as leverage rises. Expected log wealth retains this penalty in Equation~\eqref{eq:expectedlog}. Both can peak well below the wipeout threshold (i.e., the threshold where the LETF will be forced to close, or when you get margin called), while arithmetic expected wealth keeps asking for more.

\subsection{A benchmark that rewards noise}

Imagine that the prices we observe are ``noisy'' in the sense that a series of small, random supply or demand imbalances cause each closing price to deviate from its ``natural'' price. What does this do to the arithmetic mean return? Perhaps surprisingly, it increases it in expectation! When an upward price shock unwinds, the percentage loss is smaller than the gain that preceded it. When a downward shock unwinds, the percentage gain is larger than the preceding loss. Even if the noise averages to zero in \textit{prices}, it will not average to zero in percentage returns. For small percentage noise with standard deviation \(s\), the arithmetic mean over a long sample rises by approximately \((1+\bar r)s^2\), without increasing expected terminal stock wealth (see Appendix~\ref{app:temporary_prices}). The compounded sample-mean benchmark rises for the stock and even more for the LETF. Meanwhile, the extra squared daily returns increase volatility drag, eating into actual LETF wealth.

To give an example, with \(2\%\) true daily volatility and \(20\) basis points of independent closing-price noise, the stock's compounded sample-mean benchmark rises by roughly \(0.10\%\) per year and the \(2\times\) LETF's by \(0.20\%\), while the additional volatility drag tends to reduce actual LETF wealth by approximately \(0.20\%\) per year. This concern isn't hypothetical: bid-ask effects, tick sizes, and market-on-close imbalances are ordinary features of daily trading that make ``noisy'' prices a reality. For the underlying asset, intermediate price noise cancels out of cumulative log returns and expected log return, while noise in the terminal price remains because it changes the wealth actually realized.

\subsection{Sampling and modeling choices can change the benchmark}

Compounding daily and weekly sample arithmetic means can produce different benchmarks, and even opposite rankings, for the same investments over the same holding period. This is all despite the fact that actual terminal wealth has not changed. Of course, this issue goes beyond sampling, in the sense that even if we knew the entire distribution from which daily returns were drawn, we would not know expected terminal wealth because we must also specify how returns are linked across periods. Drawing days independently lets the model mix and match historical returns. Drawing whole blocks keeps the returns inside each block together. That changes which paths the model can generate and how likely they are, and therefore expected terminal wealth. Expected cumulative log returns require no such assumption on the dependence of returns: logs add through time, and expectations of sums do not require independence.

While many discussions of LETFs focus on the effects of autocorrelation on terminal wealth, we would like to highlight that the effects of serial dependence on compound returns are not unique to LETFs.  In fact, serial dependence also affects expected terminal wealth when an investor holds an ordinary stock and never rebalances. To see this, let $r_1$ and $r_2$ be the stock's simple returns over two periods, each with arithmetic mean $\mu$. Expected terminal wealth per dollar invested is
\begin{equation}
\mathbb{E}[G_2]
=
\mathbb{E}[(1+r_1)(1+r_2)]
=
(1+\mu)^2+\operatorname{Cov}(r_1,r_2).
\end{equation}
Holding the one-period return distributions fixed, positive serial covariance raises expected terminal wealth relative to independent returns, while negative serial covariance lowers it. In other words, no rebalancing is required for this covariance term to appear in expected terminal wealth.

For a costless LETF that resets leverage between the two periods, the corresponding expression is
\begin{equation}
\label{eq:autocorr}
\mathbb{E}[G_{L,2}]
=
\mathbb{E}[(1+Lr_1)(1+Lr_2)]
=
(1+L\mu)^2+L^2\operatorname{Cov}(r_1,r_2).
\end{equation}
Equation \ref{eq:autocorr} shows that leverage changes the \textit{magnitude} of the covariance term, but its presence is a generic property of compounded returns. 
Serial correlation can still affect the difference in expected returns between rebalancing and levering once\citep{Bessembinder2026Volatility}.

Over longer horizons, expected wealth also depends on cross-products involving more than two dates. Drawing whole blocks preserves the historical sequence of returns within each block, including any continuation or reversal, while drawing individual days independently from the distribution of past returns can break up those sequences. This choice of how to sample the distribution of realized returns can therefore change expected terminal wealth even when the models use the same historical daily returns. 

Table~\ref{tab:leverage_process_sensitivity} illustrates this by comparing models that draw individual days, whole years, or twenty-year blocks from a century of market returns. At \(3\times\), changing what we draw moves expected twenty-year wealth from \$1.77 million to \$27.05 million. Expected cumulative log returns stay the same within each panel. In all models leverage resets daily. Appendix~\ref{app:payoff_distribution} gives the unlevered comparison with additional draw intervals. Appendix~\ref{app:block_estimation_bias} and Table~\ref{tab:iid_estimation_bias} show how block length also affects estimation bias across random calibration samples.

\begin{table}[htbp]
\centering
\small
\setlength{\tabcolsep}{4pt}
\begin{tabular}{@{}lrrrrr@{}}
\toprule
Daily leverage
& \shortstack{Daily draws\\\(\E[W_{L,20}]\)}
& \shortstack{Annual blocks\\\(\E[W_{L,20}]\)}
& \shortstack{Two-decade blocks\\\(\E[W_{L,20}]\)}
& \shortstack{Annual /\\two-decade}
& \shortstack{All three models\\within each panel\\
\(\exp\{\E[\log W_{L,20}]\}\)} \\
\midrule
\multicolumn{6}{@{}l}{
\textit{January--December years: January 1926--December 2025}
} \\[5pt]
\(1\times\)
& \$100,526
& \$99,971
& \$79,058
& \(1.26\times\)
& \$71,249 \\[5pt]
\(2\times\)
& \$1,011,005
& \$996,170
& \$465,450
& \(2.14\times\)
& \$253,835 \\[5pt]
\(3\times\)
& \$10,172,455
& \$10,027,175
& \$2,383,337
& \(4.21\times\)
& \$443,858 \\
\midrule
\multicolumn{6}{@{}l}{
\textit{July--June years: July 1926--June 2026}
} \\[5pt]
\(1\times\)
& \$102,204
& \$117,470
& \$77,270
& \(1.52\times\)
& \$72,444 \\[5pt]
\(2\times\)
& \$1,045,038
& \$1,727,477
& \$408,591
& \(4.23\times\)
& \$262,459 \\[5pt]
\(3\times\)
& \$10,690,484
& \$27,050,302
& \$1,770,639
& \(15.28\times\)
& \$466,770 \\
\bottomrule
\end{tabular}
\caption{Twenty-year projections for a \$10,000 investment. Models draw historical days, years, or twenty-year blocks with replacement. Leverage resets daily in every model. Each panel uses 100 complete years, grouped January--December or July--June. Twenty-year blocks are contiguous and nonoverlapping, beginning at the start of each calibration window, giving five historical blocks per panel. Both the annual-block boundaries and the calibration window change between panels. The daily model draws each projected year's observation count from the historical annual counts, then independently draws that many daily returns. The ratio divides annual-block by two-decade expected wealth. Expected cumulative log returns agree across the three models within each panel. Results are exact expectations, not Monte Carlo estimates. Strategies exclude fees, financing costs, and intraday termination provisions. Data are daily total returns, including dividends, from the CRSP Value-Weighted Index of the S\&P 500 Universe (index 1000500). Appendix~\ref{app:payoff_distribution} provides further details.}
\label{tab:leverage_process_sensitivity}
\end{table}

There is a broader issue with the terminology here, which we believe does the average investor no favors. ``Expected return'' sounds like the answer to a simple question: ``What can I reasonably expect to end up with by investing in this product?'' But arithmetic average returns do not measure how invested money is expected to compound over time.

Meanwhile, ``expected log wealth'' sounds like something no one cares about. For example, when you're reading the prospectuses of LETFs, I doubt you'll see anything about expected log wealth. Despite the fact that it sounds like an artifact of an investor's utility function, we believe that using it to describe compound growth requires no assumption of log utility. Under stationary, ergodic returns, the long-run compound growth rate per period is\footnote{This statement requires positive gross returns and \(\E[|\log(1+r_t)|]<\infty\). Ergodicity makes the time average of log returns converge almost surely to its expectation. Independence is unnecessary, but the earlier comparisons of dependence structures hold the one-period return distributions fixed. Without ergodicity, unconditional expected log return need not equal the growth rate along each realized path.}
\begin{equation}
g_{\mathrm{compound}}
=\exp\!\left\{\E[\log(1+r_t)]\right\}-1.
\end{equation}

\subsection{Trading costs and the closing price}

For a conventional stock purchase, there is a simple way to think about trading costs: the stock was quoted at \$50, but the fund bought it at \$50.20. The trading cost, also known as execution shortfall, was 20 cents.

Equity LETFs work differently. They commonly rebalance by increasing or decreasing swap exposure with banks and other dealers at the closing price. The counterparties hedge these adjustments by trading in the closing auction. When those hedges execute at the official close, the execution price and the reference price are identical. Price impact, however, is still there, but it instead appears as a change in the closing price relative to the close that would have occurred without the rebalance. In other words, the price impact is baked into the closing price and the issue is that we do not observe what that price would have been without the LETF rebalance.

More broadly, price impact changes the underlying's path of returns. The wealth identity in Equation~\eqref{eq:2xidentity} already uses that observed path, so subtracting the same effect again would count it twice. Absent investor flows, long leveraged and inverse funds buy after gains and sell after losses. Their hedging can amplify closing-price moves, raising \(\sum_t r_t^2\) and therefore the leading volatility-drag term in Equation~\eqref{eq:costloggap}.

Of course, a LETF can exist without moving its underlying's price very much. Absent investor flows, the predicted rebalance is the underlying's daily return multiplied by \(\sum_i L_i(L_i-1)A_{i,t-1}\), summed across funds tracking that underlying. What matters is how large the resulting hedging trade is relative to available closing-auction liquidity, after flows and dealer netting. A volatile stock with little AUM in LETFs tracking it can therefore experience little impact, while a modest move with substantial LETF AUM can generate price impact.

No price reversal is required. Consider two stock-price paths:
\[
\begin{aligned}
\text{Without price impact:}&\quad \$100\to\$110\to\$112,\\
\text{With price impact:}&\quad \$100\to\$111\to\$112.
\end{aligned}
\]
Suppose closing-auction buying raises the first close from \$110 to \$111, but both paths finish at \$112. The stock rises on both days in both paths. The exact discrete-time wealth identity is
\[
G_{2,2} =\left(\frac{S_2}{S_0}\right)^2 \underbrace{\prod_{t=1}^{2} \frac{1+2r_t}{(1+r_t)^2}}_{\text{volatility-drag multiplier}}.
\]
The underlying ends at \$112 in both cases, so \((S_2/S_0)^2=1.12^2=1.2544\) for both paths. Only the volatility-drag multiplier differs, as Table~\ref{tab:closing_price_impact} shows:

\begin{table}[htbp]
\centering
\caption{Closing-price impact and terminal wealth without a price reversal.}
\label{tab:closing_price_impact}
\begin{tabular}{lrr}
\toprule
& Without price impact & With price impact \\
\midrule
Underlying cumulative return & \(12\%\) & \(12\%\) \\
LETF cumulative return & \(24.3636\%\) & \(24.1982\%\) \\
Squared reset returns, \(\sum_t r_t^2\)
& \(0.010331\) & \(0.012181\) \\
Exact drag multiplier & \(0.991419\) & \(0.990100\) \\
\bottomrule
\end{tabular}
\end{table}

That dollar of price impact on the second path adds volatility drag and cuts the LETF's return by \(0.1654\) percentage points. The stock never falls. The initial price bump simply fades while the stock keeps rising.

In this example, we can see that the stock investor ends up in the same place either way. The LETF investor, however, does not. The cost is already reflected in the realized return path, so there is no second market-impact bill to subtract.

\section{Empirical Applications}
\label{sec:applications}

\subsection{Intraday termination: the Lucid example}

A $2\times$ LETF does not need its underlying stock to \textit{close} down $50\%$ to lose everything. Swap agreements can permit intraday termination, preventing the LETF from participating in a subsequent recovery. On July 14, 2026, bankruptcy rumors that Lucid subsequently denied sent its shares down more than $50\%$ intraday. The stock recovered to close down $16.15\%$, but the GraniteShares $2\times$ Long LCID Daily ETF (LCDL) did not. Its swap counterparty closed out the position, leaving negative NAV and nothing for shareholders. In other words, the stock recovered but because the fund closed, it could not.

The risk of a LETF swap being closed out can make a bad situation worse: a dealer unwinding a long stock hedge sells into an already falling market and proprietary traders anticipating the liquidation may also short the stock, adding selling pressure. Any LETF backtest must account for intraday prices and the swap's termination terms: closeout may occur before the theoretical $50\%$ wipeout point, and liquidation itself can change the price path that triggers it. This crash risk may help explain why single-stock LETFs pay higher financing costs: the swap dealer is effectively selling deep out-of-the-money crash insurance, not just lending money, and needs to be compensated for this.

\subsection{The convexity bonus: NVIDIA}
\label{sec:nvda_convexity}

In this subsection, we use NVIDIA as an example to illustrate how large the convexity bonus can be. Table~\ref{tab:nvda_convexity} reports the decomposition for 2023 and 2024. Its $239\%$ return in 2023 produces a bonus of $571$ percentage points, larger than the $478\%$ obtained by simply doubling the stock's return. After estimated volatility drag, the costless $2\times$ return is $809\%$. In 2024, the corresponding estimate is $455\%$, versus $342\%$ from simply doubling NVIDIA's return.

No $2\times$ NVIDIA ETF had a full-calendar-year history in 2023, but in 2024 NVDX returned approximately $382\%$, compared with our costless estimate of \(455\%\). For illustration, a $1\%$ fee haircut and a further $12\%$ financing haircut give
\[
(1+4.5523)\times0.99\times0.88-1=383.7\%.
\]
Costs multiply terminal wealth rather than simply being subtracted from the cumulative return.

\begin{table}[htbp]
\centering
\small
\setlength{\tabcolsep}{4pt}
\begin{tabular}{lrrrrrrr}
\toprule
Year &
\shortstack{NVDA\\return} &
\shortstack{Daily\\vol.} &
$2R_S$ &
\shortstack{Convexity\\bonus (pp)} &
\shortstack{Estimated\\drag multiplier} &
\shortstack{Estimated\\costless return} &
\shortstack{Actual\\NVDX return} \\
\midrule
2023 & $239\%$ & $3.05\%$ & $478\%$ & $571.2$ & $0.791$ & $809.1\%$ & --- \\
2024 & $171\%$ & $3.30\%$ & $342\%$ & $292.4$ & $0.756$ & $455.2\%$ & $382.25\%$ \\
\bottomrule
\end{tabular}
\caption{NVIDIA's convexity bonus and estimated $2\times$ returns.}
\label{tab:nvda_convexity}

\vspace{0.4em}
\begin{minipage}{0.95\textwidth}
\footnotesize
\emph{Notes:} Estimated costless returns use $(1+2R_S+R_S^2)\exp(-\sum_t r_t^2)-1$. For 2023, squared returns are approximated by $252\sigma_d^2$. For 2024, their measured sum is $0.279681$. The convexity bonus is $R_S^2$ and enters before drag, which multiplies the entire gross payoff. Estimates omit higher-order discrete compounding terms, fees, and financing costs. The financing haircut illustrated in the text is not an observed cost.
\end{minipage}
\end{table}

\subsection{XIV, average daily returns, and cumulative wealth}

\begin{figure}[htbp]
\centering
\includegraphics[width=0.78\textwidth]{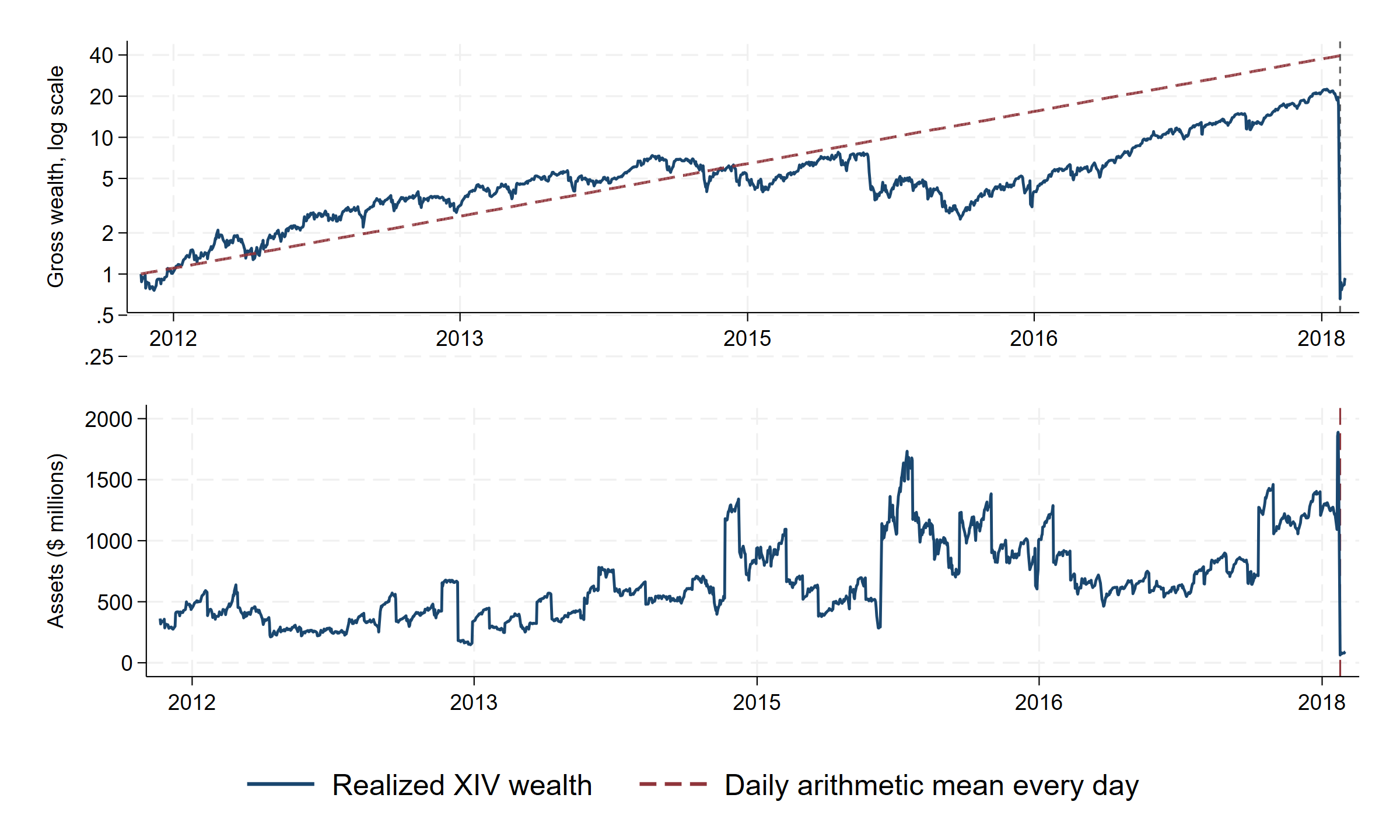}
\caption{XIV returns and investor exposure. The top panel contrasts realized cumulative wealth with the hypothetical wealth produced by earning XIV's arithmetic average return every day. The bottom panel reports month-end assets. The vertical line marks February 5, 2018. Sources: CRSP daily returns and monthly assets.}
\label{fig:xiv}
\end{figure}

XIV, an inverse VIX ETN, provides a real-world example of the pitfalls of evaluating LETF performance using arithmetic average returns. Figure~\ref{fig:xiv} shows that from November 2011 through February 15, 2018, XIV had an arithmetic average daily return of $0.234\%$. If XIV had earned exactly that return every day, one dollar would have grown to \$40.25 over the sample. Compounding its actual daily returns, however, left that dollar worth only \$0.94. This relatively small full-sample loss conceals an enormous decline near the end of XIV's life: its value fell from \$108.37 on February 2, 2018, to a redemption value of \$5.99.

The example also highlights the importance of how many dollars investors had exposed at each point in time. XIV had \$1.86 billion in assets at the end of January 2018 and lost $95.16\%$ from February 1 through February 15. Assuming no net flows during those two weeks, investors lost approximately \$1.77 billion in the collapse.

\subsection{Investor flows and timing in Korean LETFs}

\begin{figure}[htbp]
\centering
\includegraphics[width=0.78\textwidth]{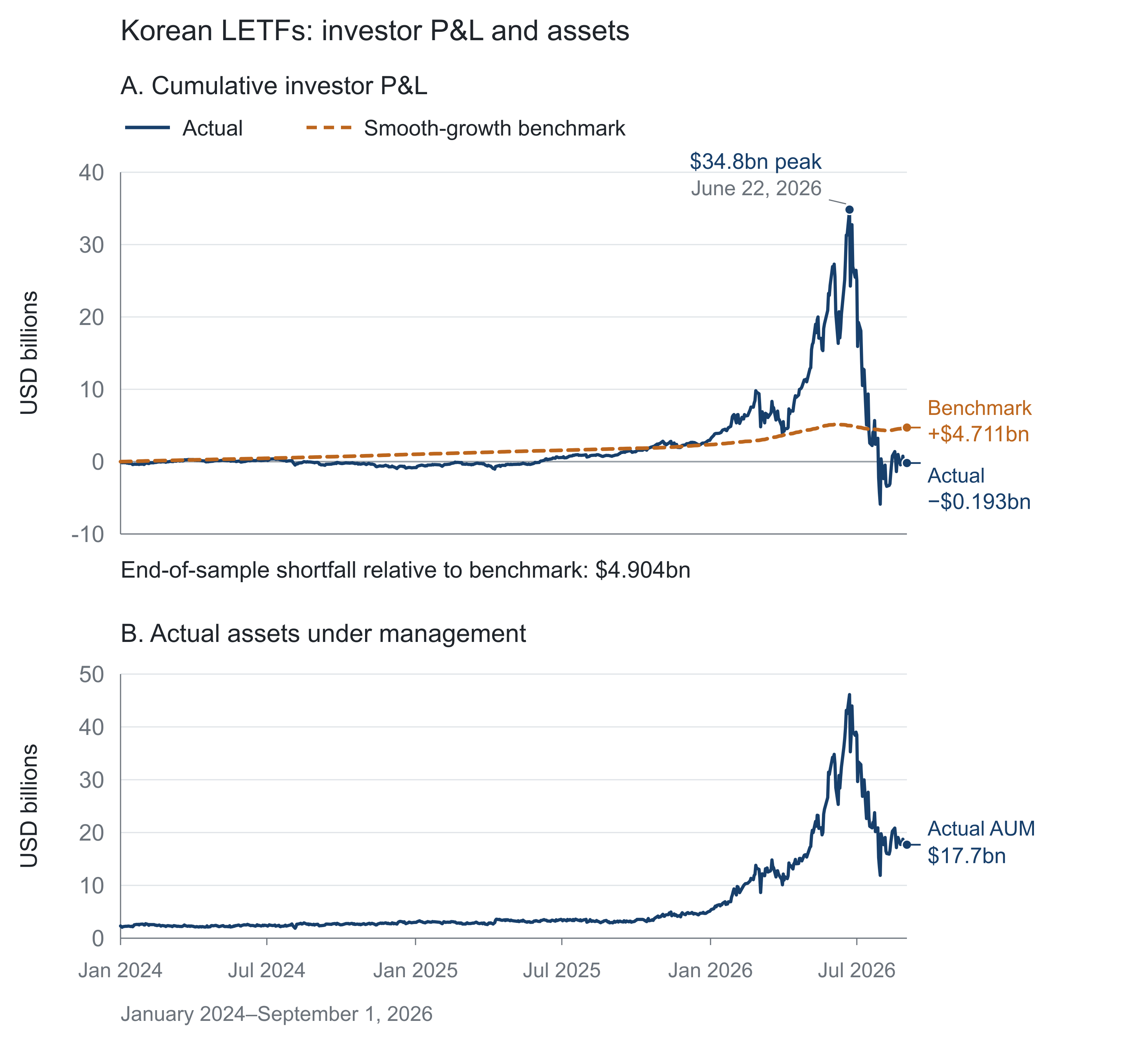}
\caption{Investor P\&L and assets in ten large leveraged funds tracking Korean assets from January 2024 through September 1, 2026. The top panel compares actual cumulative investor P\&L with a smooth-growth benchmark that preserves each fund's full-sample compound return, initial assets, dollar contributions, and redemption percentages. The bottom panel reports actual aggregate assets exposed to subsequent returns. Source: author-supplied Bloomberg data.}
\label{fig:korea}
\end{figure}

Ten LETFs tied to Korean equities illustrate how investor flows affect aggregate investor P\&L. Specifically, we focus on products providing levered exposure to Samsung Electronics, SK Hynix, the KOSPI 200, KOSDAQ 150, or Korean semiconductor indexes. Because the amount invested in these products changed substantially over the sample, fund returns alone do not reveal how many dollars investors gained or lost. We calculate investor P\&L by applying each product's daily return to its beginning-of-day assets. Figure~\ref{fig:korea} shows that cumulative investor P\&L peaked at \$34.8 billion on June 22, 2026, when the products held \$46.1 billion in assets. By September 1, cumulative investor P\&L had fallen to negative \$193 million, while assets had fallen to \$17.7 billion.

To measure the flow timing relative to smooth growth 
we first construct a ``timing-neutral'' benchmark for each product. Specifically, we replace the product's actual sequence of daily returns with a constant daily return equal to its full-sample geometric mean. Unlike the arithmetic mean, the geometric mean incorporates volatility drag. Consequently, in the absence of investor flows, the benchmark follows a smooth path while producing the same beginning-to-end compound return as the actual product. We then apply the actual sequence of flows to this smoothed path. On each day, after applying the constant return, we add the same dollar amount contributed to the actual product or redeem the same percentage of post-return assets that was redeemed from the actual product. Thus, the benchmark preserves the timing and dollar amount of contributions and the timing and proportional size of redemptions while removing daily fluctuations in returns.

We calculate benchmark P\&L as ending assets plus withdrawals, minus initial assets and subsequent contributions. Across the ten funds, this benchmark produces \$4.711 billion in investor P\&L, compared with an actual loss of \$193 million, a \$4.904 billion shortfall. This benchmark-relative timing gap measures how the observed contributions and redemption percentages interacted with the realized return sequence, compared with smooth growth delivering the same full-sample compound return.

\subsection{Hourly reset intervals}
\label{sec:hourly}

In August 2026, Defiance filed to launch sixteen $2\times$ leveraged ETFs that would reset six times during the trading day \citep{Lee2026Hourly,SEC2026HourlyReset}. The number of resets does not by itself determine funding costs, which depend on how much the fund borrows over time. More frequent resets can, however, increase turnover and trading costs. The proposed indexes finance at the Secured Overnight Financing Rate (SOFR) and charge one basis point on each change in notional exposure during the six execution periods and at the close. Appendix~\ref{app:intraday_mechanics} describes the observation, execution, and closing-price windows. A long-run evaluation should therefore use reset-to-reset returns and the trading costs generated by the realized rebalancing path.

Daily-reset LETFs benefit from a trading business built specifically around liquidity at the close. The closing auction coordinates liquidity and establishes the official price used to measure the fund's benchmark. Banks also offer guaranteed-close trades, under which they commit to transact with the fund at that official price and assume the risk of executing the hedge. If a fund's swap is resized at the same closing price used to calculate its daily return, there is no execution-price mismatch on the reset itself. An hourly-reset fund would effectively ask banks to reproduce the guaranteed-close business six times during the trading day.

Scheduled intraday rebalancing may also create additional volatility in the underlying stocks. Absent fund flows, daily-reset LETFs trade in the direction of the underlying stock's return. In the sample studied by~\citet{ivanov2018leveraged}, flows offset as much as $85\%$ of the predicted rebalancing, and the estimated effects on late-day returns and volatility are economically small. Hourly-reset funds would move this potential trading from one period near the close to six published windows during the day.

The same mechanism could arise with hourly-reset ETFs. Once traders observe a fund's assets and the underlying stock's return, they can estimate whether the fund or its swap counterparty will need to buy or sell at the next reset. Because the reset schedule and time-weighted average price (TWAP) windows are published in advance, traders can buy before an expected purchase or sell before an expected sale. This front-running can move prices against the fund before its rebalance occurs.

Anticipatory trading can affect both the execution price and, by changing the price used to calculate required exposure, the size or direction of the rebalance itself. These effects should be larger when the fund is large relative to trading volume or the underlying stock is illiquid.

\section{Conclusion}

Our main point in writing this note is that long-horizon investors ultimately care about how much their invested dollars grew. And therefore, any measure of investment performance should remain tied to the growth of dollars in a given investment. Expected terminal wealth may be useful for studying average outcomes across the hypothetical return histories generated by a statistical model, or for evaluating one-off bets, but it does not measure the rate at which wealth compounds in a given product over time.

One way to estimate expected terminal wealth under an iid return model is the compounded sample-mean benchmark, typically implemented by replacing the unknown population mean with the arithmetic average return from a single realized time series and compounding that return over the holding period. Estimation error, however, is not the fundamental problem with the compounded sample-mean benchmark: even with the population mean known exactly, expected terminal wealth would generally not describe long-run compound growth. Moreover, we show that the compounded sample-mean benchmark implies strictly greater terminal wealth than the realized path whenever returns are literally not constant across periods. Leverage makes this issue even larger, as a costless \(2\times\) fund doubles daily returns but quadruples their variance.

This is not the only issue with using the compounded sample-mean benchmark to evaluate LETF performance. Common sense suggests that sampling the same return series daily, weekly, or annually should not change our conclusion about the underlying asset's long-horizon performance. Despite this, we show that the compounded sample-mean benchmark is affected by sampling frequency. Further, even when the daily return distribution is held fixed, different assumptions about whether high or low returns tend to be followed by high or low returns change the cross-products of returns across dates and can therefore change expected terminal wealth dramatically. These assumptions do not, however, change expected cumulative log returns or the long-run rate at which invested dollars are expected to compound. Finally, we show that temporary price noise can also make the compounded sample-mean benchmark look better while reducing expected LETF wealth.

All of these examples are designed to show why compounding an arithmetic average can give a misleading picture of the wealth earned by an investor in that product. The benchmark can rank investments backwards, report a large positive return for an investment that lost money, and recommend infinite leverage while realized wealth approaches zero. Long-horizon expected wealth can also be driven by economically infeasible payouts. We argue that exponentiated cumulative log returns provide a direct measure of realized compound performance in a given product. And further, we show that expected log returns avoid the arithmetic-aggregation distortions described above. This follows from how money compounds, not an assumption that investors have log utility.

Serial covariance can affect expected terminal wealth through compounding in both LETFs and ordinary stocks. Covariances between returns far apart in time can matter, not just between consecutive days. For otherwise identical joint return processes, the effect is the same whether those returns come from a stock or an LETF.

Zooming out, in \citet{MurraySammon2026} we are interested in LETFs as investment products. In this note, we argue that clear performance measures are needed to separate the effects of product design (i.e., leverage factors, reset intervals), costs (i.e., fees and funding spreads), and investor behavior (i.e., the timing of flows). Volatility drag reduces the payoff, but the less discussed convexity bonus can increase it, especially when the underlyer's cumulative return is large in either direction. The collapse of XIV shows how a fund can have positive average daily returns but a negative cumulative return over the full sample. Lucid shows why closing prices alone can miss an intraday wipeout, and hourly resets raise further questions about trading costs and execution at predictable times.

Our evidence from Korean LETFs shows that positive cumulative \textit{fund} returns do not necessarily mean that \textit{investors} made money in those products. Answering that question requires investor P\&L, which accounts for how much money investors put into each product and when they invested it. Appendix~\ref{app:clientele} discusses how product selection and changes in investor clientele affect the interpretation. Together, these examples illustrate why evaluating LETFs requires starting with the wealth earned by the dollars investors actually put at risk.

\clearpage
\appendix

\section{Which Return-Generating Process Are We Assuming?}
\label{app:payoff_distribution}

Earlier, we noted that maximizing expected payoff can be reasonable for a one-shot bet but does not measure compound growth. This raises an apparent contradiction: could an investor with a decades-long horizon simply view terminal wealth as the payoff of a one-shot bet? They can, if they believe the payoff distribution. The question is where that distribution comes from.

Take the same historical returns and build a model that independently draws days, months, years, or twenty-year blocks. These choices can give very different \(\E[W_{L,T}]\) for the same twenty-year investment horizon. Which distribution should the investor believe? The historical data do not choose the draw interval for us.

The issue goes beyond sampling. Are we drawing fresh returns each day, or do they arrive in longer blocks? That is epistemic uncertainty about which dice the market rolls, not how we choose to display the results. Holding the daily return distribution fixed, these choices can change \(\E[W_{L,T}]\), but not \(\E[\log W_{L,T}]\): logs add through time, accurately describe the rate at which wealth compounds, and expectations of their sums do not require independence. Generating \(\E[W_{L,T}]\) but using the historical geometric mean instead also makes the benchmarks agree across frequencies, because it builds in the variance penalty that the arithmetic mean leaves out. Compounding the historical geometric mean gets you the same answer as projecting log returns and exponentiating. Same calculation, different packaging.

Table~\ref{tab:block_generating_processes} starts with a \$10,000 initial investment and uses a century of market returns. In the July--June panel, change how the model draws returns and expected ending wealth moves from \$77,270 to \$117,470. Same asset, same twenty-year horizon. The geometric-mean wealth projection does not budge from \$72,444. And this is before adding leverage. Table~\ref{tab:leverage_process_sensitivity} in the main text shows what happens when we turn that dial.

Daily resampling can repeatedly draw winners or losers. But those streaks do not cancel in dollars: losses are bounded by the initial investment, while repeated gains can multiply wealth many times over and pull the average upward.

\begin{table}[htbp]
\centering
\small
\setlength{\tabcolsep}{3pt}
\begin{tabular}{@{}lrrrrr@{}}
\toprule
Draw interval
& \shortstack{Historical\\observations}
& \shortstack{Mean per draw\\\(\mu\)}
& \shortstack{Volatility per draw\\\(\sigma\)}
& \(\E[W_{20}]\)
& \(\exp\{\E[\log W_{20}]\}\) \\
\midrule
\multicolumn{6}{@{}l}{
\textit{January--December years: January 1926--December 2025}
} \\[5pt]
Daily
& 26,300
& \(0.0439\%\)
& \(1.1423\%\)
& \$100,526
& \$71,249 \\[5pt]
Monthly
& 1,200
& \(0.9652\%\)
& \(5.3753\%\)
& \$100,281
& \$71,249 \\[5pt]
Annual
& 100
& \(12.2002\%\)
& \(19.5113\%\)
& \$99,971
& \$71,249 \\[5pt]
Decade
& 10
& \(188.2331\%\)
& \(107.9248\%\)
& \$83,078
& \$71,249 \\[5pt]
Two decades
& 5
& \(690.5807\%\)
& \(339.7025\%\)
& \$79,058
& \$71,249 \\
\midrule
\multicolumn{6}{@{}l}{
\textit{July--June years: July 1926--June 2026}
} \\[5pt]
Daily
& 26,274
& \(0.0442\%\)
& \(1.1428\%\)
& \$102,204
& \$72,444 \\[5pt]
Monthly
& 1,200
& \(0.9724\%\)
& \(5.3796\%\)
& \$102,013
& \$72,444 \\[5pt]
Annual
& 100
& \(13.1088\%\)
& \(25.1837\%\)
& \$117,470
& \$72,444 \\[5pt]
Decade
& 10
& \(186.5723\%\)
& \(101.7424\%\)
& \$82,124
& \$72,444 \\[5pt]
Two decades
& 5
& \(672.6980\%\)
& \(251.2389\%\)
& \$77,270
& \$72,444 \\
\bottomrule
\end{tabular}
\caption{Twenty-year projections for a \$10,000 investment. The panels use January 1926--December 2025 and July 1926--June 2026, so both the sample endpoints and the annual grouping change. We draw historical days or whole blocks, allowing repeats: 240 months, twenty years, two decades, or one twenty-year block. Historical blocks do not overlap and begin at the start of each sample, and each block's return compounds the daily returns inside it. For daily draws, we first pick a historical year's trading-day count, then independently pick that many days from the full sample. We repeat this for twenty projected years. The reported means and volatilities are per draw, not annualized. These are exact model averages, not Monte Carlo estimates. Expected cumulative log returns agree within each panel. The twenty-year-block model has just five historical observations. Source: daily returns, including dividends, from the CRSP Value-Weighted Index of the S\&P 500 Universe (index 1000500).}
\label{tab:block_generating_processes}
\end{table}

Table~\ref{tab:leverage_process_sensitivity} shows how leverage magnifies the effect of the assumed generating process on expected wealth.

\subsection{Effect of Block Length on Compounding an Estimated Mean}
\label{app:block_estimation_bias}

Here we'll look more generally at how upward bias in the compounded sample mean benchmark varies with the block-length we use to compute the sample mean. We'll motivate this by assuming daily log returns are iid normal with mean \(0.0004\) and standard deviation \(3\%\). Next we draw forty years of returns from this distribution and use this sample to forecast wealth twenty years ahead, using a \$10,000 starting investment. For each block length \(k\), we compound the daily returns within each nonoverlapping block, calculate the sample arithmetic mean block return \(\bar r_k\), and forecast terminal wealth as \(\widehat W_H(k)=\$10{,}000(1+\bar r_k)^{5040/k}\): the compounded sample mean benchmark with block length $k$. This forecast is fixed once we draw a sample, but the samples themselves are random. We want to measure how biased this forecast is. The true expected terminal wealth is \$725,300.

Table~\ref{tab:iid_estimation_bias} averages these forecasts across possible forty-year histories. The results are not from Monte Carlo: they can be computed exactly. The daily benchmark averages \$2.25 million, while the annual benchmark averages \$2.45 million. The true return-generating process is identical in every row: only the frequency used to estimate its mean changes. Longer blocks do not consistently reduce the bias, but the bias disappears when the block length equals the forecast period. The forecast of expected log wealth is unbiased.

\begin{table}[htbp]
\centering
\small
\setlength{\tabcolsep}{3pt}
\begin{tabular}{@{}lrrrrr@{}}
\toprule
Draw interval
& \shortstack{Historical\\observations}
& \shortstack{Expected\\arithmetic\\benchmark}
& \shortstack{Arithmetic\\benchmark\\bias}
& \shortstack{Arithmetic\\relative\\bias}
& \shortstack{Log\\forecast\\bias} \\
\midrule
Daily
& 10,080 & \$2,254,922 & \$1,529,622
& \(210.9\%\) & \(0\) \\
Monthly
& 480 & \$2,267,982 & \$1,542,682
& \(212.7\%\) & \(0\) \\
Annual
& 40 & \$2,452,525 & \$1,727,225
& \(238.1\%\) & \(0\) \\
Decade
& 4 & \$2,295,585 & \$1,570,285
& \(216.5\%\) & \(0\) \\
Twenty-year
& 2 & \$725,300 & \$0
& \(0.0\%\) & \(0\) \\
\bottomrule
\end{tabular}
\caption{Twenty-year forecasts for a \$10,000 investment, estimated from forty years of iid daily log returns with mean \(0.0004\) and standard deviation \(3\%\). Years contain 252 trading days and monthly blocks contain 21 trading days. Arithmetic benchmark bias is measured against true expected terminal wealth, \$725,300. The estimated expected log wealth is \(\widehat m_H=\log W_0+(H/C)\sum_{t=1}^{C}\log(1+r_t)\), where \(C=10{,}080\) and \(H=5{,}040\). It is unbiased and unchanged by grouping the same returns into blocks. Results use exact lognormal moments, not Monte Carlo estimates.}
\label{tab:iid_estimation_bias}
\end{table}

Let \(C\) be the number of calibration days, \(H\) the forecast horizon in days, and \(k\) the block length. Assume both \(C\) and \(H\) are divisible by \(k\), and define
\[
n=\frac{C}{k}, \qquad p=\frac{H}{k}.
\]
Suppose daily log returns \(\ell_t\equiv\log(1+r_t)\) are iid \(N(b,\sigma^2)\). Each historical block has gross return
\[
G_{j,k} =\prod_{t=(j-1)k+1}^{jk}(1+r_t), \qquad j=1,\ldots,n.
\]
The compounded sample-mean forecast is
\begin{equation}
\widehat W_H(k)
=
W_0\left(\frac{1}{n}\sum_{j=1}^{n}G_{j,k}\right)^p.
\end{equation}
True expected terminal wealth is
\begin{equation}
M_H\equiv\E[W_H]
=W_0\exp\left[H\left(b+\frac{\sigma^2}{2}\right)\right].
\end{equation}
The forecast's exact expectation is
\begin{equation}
\E[\widehat W_H(k)]
=
\frac{M_H}{n^p}
\sum_{\substack{a_1,\ldots,a_n\geq0\\
a_1+\cdots+a_n=p}}
\frac{p!}{a_1!\cdots a_n!}
\exp\left[
\frac{k\sigma^2}{2}
\sum_{j=1}^{n}a_j(a_j-1)
\right].
\label{eq:block_estimation_bias}
\end{equation}
Thus, expected bias is \(\E[\widehat W_H(k)]-M_H\). When the block length equals the forecast horizon, \(p=1\), the exponential terms all equal one and the forecast is unbiased. When \(p>1\) and \(\sigma>0\), the bias is strictly positive.

\section{Discrete-Time Leveraged Return Identity}
\label{app:discrete_identity}

Let \(r_t\) denote the underlying return over reset interval \(t\). Ignoring fees and financing costs, a \(2\times\) LETF satisfies
\begin{equation}
G_{2,T}
=
\prod_{t=1}^{T}(1+2r_t)
=
\left(\frac{S_T}{S_0}\right)^2
\prod_{t=1}^{T}\frac{1+2r_t}{(1+r_t)^2}.
\end{equation}

For small reset-period returns,
\begin{equation}
G_{2,T}
\approx
\left(\frac{S_T}{S_0}\right)^2
\exp\left(-\sum_{t=1}^{T}r_t^2\right).
\end{equation}

More generally, for leverage \(L\),
\begin{equation}
G_{L,T}
=
\left(\frac{S_T}{S_0}\right)^L
\prod_{t=1}^{T}
\frac{1+Lr_t}{(1+r_t)^L},
\end{equation}
which, for small returns, gives
\begin{equation}
G_{L,T}
\approx
\left(\frac{S_T}{S_0}\right)^L
\exp\left[
-\frac{L(L-1)}{2}\sum_{t=1}^{T}r_t^2
\right].
\end{equation}

\section{Arithmetic Returns and Temporary Price Movements}
\label{app:temporary_prices}

Take any positive price path \(P_0,\ldots,P_T\), and add independent price multipliers \(X_t>0\), with \(\E[X_t]=1\) and \(X_0=1\). Assume the noise is bounded and small enough to keep all leveraged gross returns positive. Expected terminal stock wealth is unchanged:
\[
\E\left[\frac{P_TX_T}{P_0}\right]=\frac{P_T}{P_0}.
\]

Consider one noisy intermediate price, holding everything else fixed. Its multiplier \(x\) changes the two adjacent gross returns to \(ax\) and \(b/x\), for constants \(a,b>0\). Their sum is convex in \(x\). Jensen's inequality therefore shows that mean-zero price noise raises the expected arithmetic average return. Leverage multiplies this increase by \(L\). The compounded sample-mean benchmarks also rise in expectation because they are increasing, convex functions of that average.

For an \(L>1\) strategy, those same two returns contribute the following factor to terminal wealth:
\[
F(x)
=(1-L+Lax)(1-L+Lb/x),
\qquad
F''(x)=\frac{2L(1-L)b}{x^3}<0.
\]
Thus terminal LETF wealth is concave in each intermediate price multiplier. Its logarithm is also concave wherever wealth is positive. Applying Jensen's inequality one multiplier at a time shows that the noise lowers both expected LETF wealth and expected log wealth. Independent terminal-price noise enters LETF wealth linearly, so it does not change expected wealth and cannot increase expected log wealth.

The inequalities are strict when intermediate noise is nondegenerate. They concern expectations over the added noise, not a claim that every noisy path performs worse.

\section{Intraday Reset LETF Mechanics}
\label{app:intraday_mechanics}

The six resets are scheduled roughly one hour apart. Before each reset, the index measures the stock's average price over a short observation window. Specifically, it records the last-traded price once per second and takes the arithmetic average of those observations, typically called a time-weighted average price (TWAP). The six observation windows occur between 9:35 a.m. and 2:35 p.m., although the filings do not report the exact start and end of each window. The index uses each observation TWAP to calculate the new position required to restore $2\times$ exposure.

A separate execution window follows each observation window. For purposes of calculating the index (i.e., what the hourly reset LETFs would be benchmarked against and expected to track), the index assumes that the required increase or decrease in exposure occurs at the TWAP during that execution window. The six execution windows are scheduled between 9:55 a.m. and 3:00 p.m. The fund would attempt to adjust its swaps, options, or stock positions during these periods, but its actual trading prices may differ from the execution TWAP used by the index, potentially creating tracking error.

After the sixth reset, a final TWAP from 3:30 to 3:35 p.m. determines the index's end-of-day value. This is a closing reference price, not a seventh reset. The fund's NAV is calculated at 4:00 p.m. Price changes between 3:35 and 4:00 therefore affect the fund's NAV but not the index's closing value. The fund can differ from the index for this reason and because its actual reset trades may occur at prices different from the execution TWAPs.

It also essentially creates a game of chicken among the traders attempting to front-run the reset. Suppose traders expect the fund to buy and position themselves by buying first. If another investor then sells aggressively and pushes the stock price down, the fund's required purchase may become smaller or even turn into a sale. The traders who bought in anticipation of the reset must then decide whether to hold their positions or sell before everyone else does. Their attempts to exit can push the price down further and change the fund's required trade again.

Trading in Coca-Cola and IBM on July 19, 2012 illustrates the risk created by a predictable trading schedule. \citet{Cochrane2012Weird} documented price jumps on the hour while the trading was underway and later connected them to a time-weighted trading algorithm. A subsequent study found synchronized half-hour reversals in Coca-Cola, McDonald's, IBM, and Apple and attributed them to repeated automated hedging of large derivatives positions \citep{LehalleLasnier2012Sawtooth}. The trading temporarily moved prices in one direction and prices reversed after the orders stopped.

\section{The Continuous-Time Return Identity}
\label{app:continuous_identity}

Let \(S_t\) be a positive continuous semimartingale. Define
\[
\mathrm{RV}_{S,T}\equiv[\log S]_T
=
\lim_{\max_i(t_i-t_{i-1})\to0}
\sum_i
\left[
\log\left(\frac{S_{t_i}}{S_{t_{i-1}}}\right)
\right]^2,
\]
where \(0=t_0<t_1<\cdots<t_n=T\) is a partition of the holding period and the limit is in probability. Put simply, divide the holding period into shorter and shorter intervals, calculate the log return in each, and add the squared returns. Quadratic variation is the limit of that sum.

Ignoring fees and financing costs, a self-financing strategy maintaining leverage \(L\) satisfies \(\frac{dW_{L,t}}{W_{L,t}}=\allowbreak L\,\frac{dS_t}{S_t}\). It\^{o}'s formula gives
\[
d\log W_{L,t}
=
L\,d\log S_t
-\frac{L(L-1)}{2}\,d[\log S]_t.
\]
Integrating and exponentiating yields
\[
G_{L,T}
=
\frac{W_{L,T}}{W_{L,0}}
=
\left(\frac{S_T}{S_0}\right)^L
\exp\left[-\frac{L(L-1)}{2}[\log S]_T\right].
\]
Setting \(L=2\) gives Equation~\eqref{eq:2xidentity}. No GBM or independence assumption is required.

\section{Additional Applications}
\label{app:additional_applications}

\subsection{Fees and financing over long horizons}
\label{app:long_run_costs}

In \cite{MurraySammon2026} paper, we examined broad-equity LETFs during a period with strong equity returns. Extending the history to include the 2000--2002 tech bust changes two things at once. First, it adds a major draw-down.  Second, it adds many years of recurring expenses.

Specifically, we examine ULPIX and UOPIX. They are slightly different than what we examine in the rest of the paper, in that they are leveraged open-end mutual funds rather than ETFs. ULPIX targets twice the daily return of the S\&P 500, while UOPIX targets twice the daily return of the Nasdaq-100.  Despite being a bit different, these are the longest running equivalents that were run in real time that we could find in the data.

We obtain fund returns and historical expense ratios for ULPIX and UOPIX from CRSP mutual fund data. The fees on these funds are large.  For example, ULPIX currently reports a $1.47\%$ expense ratio.  If that annual fee were charged for 25 years, deducting one-twelfth of it each month would leave about $69\%$ of the wealth generated by an otherwise identical gross return path.

\begin{figure}[H]
\centering
\includegraphics[width=0.97\textwidth]{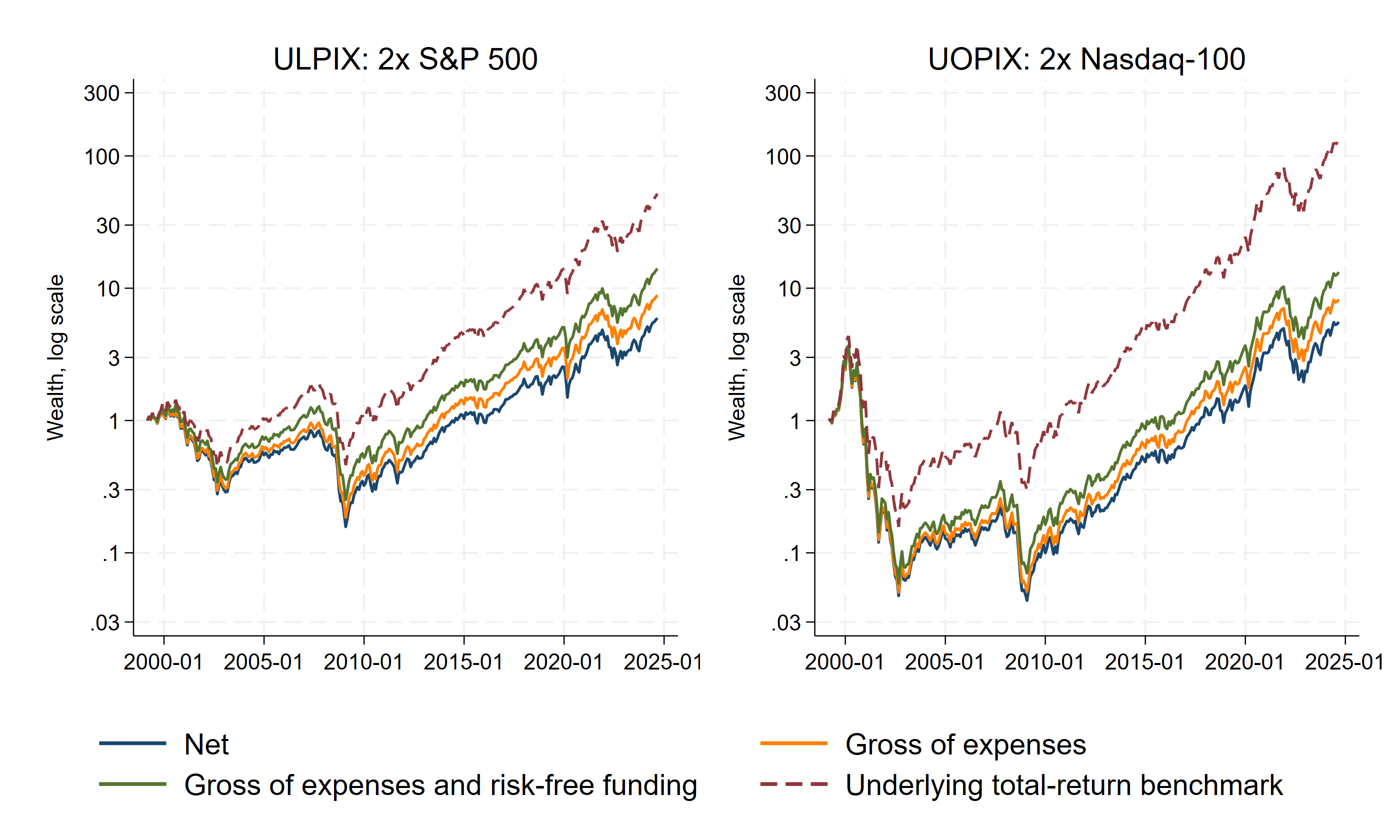}
\caption{ULPIX and UOPIX net returns, cost adjustments, and underlying-index benchmark. All paths include reinvested distributions and start at one dollar on March 31, 1999, with monthly returns from April 1999 through September 2024. The gross-of-expenses monthly return equals the CRSP net total return plus the contemporaneous annual expense ratio divided by 12. The next path also adds the compounded monthly Fama--French risk-free return, a conservative estimate of the base funding cost for one dollar of additional exposure. The dashed path exponentiates the leverage ratio times the underlying total-return index's cumulative log return, calculated as $\exp\{\sum_t 2\log(1+r_{u,t})\}$, where $r_{u,t}$ includes reinvested dividends. It is not a daily-reset investable strategy. The common start date reflects the availability of the Nasdaq-100 total-return index. Sources: CRSP, Kenneth French's data library, S\&P 500 Total Return via Yahoo Finance, and Nasdaq-100 Total Return via FRED.}
\label{fig:longruncosts}
\end{figure}

The next step is to separate the effect of a long-run investment in a daily-reset LETF from the effect of long run compounding on fees and leverage.   Specifically, Figure~\ref{fig:longruncosts} separates expenses from a conservative estimate of financing costs -- the risk free rate. The net line is the reported CRSP total return (i.e., the return net of fees and financing costs), including reinvested dividends and capital-gain distributions. The gross-of-expenses line adds one-twelfth of the contemporaneous annual expense ratio to each monthly return. The next line also adds the monthly Fama--French risk-free return i.e., adds back estimated funding costs. The dashed line compounds twice the underlying total-return index's monthly log return, including reinvested dividends, and therefore shows the underlying index times the leverage ratio, before volatility drag and implementation costs.\footnote{Note, the dashed line is \textit{not} the same thing as a ``lever once'' strategy. As we discuss in the main body of the paper, a position levered once to $2\times$ and then held without resetting becomes nearly unlevered after a large increase in the underlying asset's value. Ignoring financing costs, such positions would end this sample with leverage of approximately $1.07\times$ for the S\&P 500 and $1.05\times$ for the Nasdaq-100. The dashed lines in Figure~\ref{fig:longruncosts} use a different comparison: they square the underlying total-return wealth multiple. They are hypothetical before-drag benchmarks, rather than positions whose leverage drifts down over time.} 

Figure~\ref{fig:longruncosts} shows that, from March 31, 1999 through September 30, 2024, reported expenses reduced terminal wealth by approximately $32.8\%$ for ULPIX and $32.4\%$ for UOPIX relative to the expense-adjusted paths. Risk-free funding costs would have reduced terminal wealth by a further $37.1\%$ and $37.8\%$, respectively. These reductions apply multiplicatively. But, even after both adjustments, terminal wealth remains $72.7\%$ below the underlying-times-leverage benchmark for ULPIX and $89.9\%$ below it for UOPIX. This remaining gap includes volatility drag, funding spreads, trading costs, and other tracking differences.

Financing is another distinct cost. To measure it over a longer history, we simulate daily-reset exposure to the broad U.S. equity market beginning on August 31, 1976, when Vanguard launched the first retail index mutual fund \citep{Vanguard2026Indexing}. The daily return at leverage $L$ is
\begin{equation}
r^{L,f}_t=Lr_t-(L-1)r_{f,t}-\frac{f}{252},
\label{eq:financingfees}
\end{equation}
where $r_t$ is the Fama--French total market return, $\mathrm{MKT-RF}+\mathrm{RF}$, $r_{f,t}$ is the daily risk-free rate, and $f$ is the annual expense ratio. A $2\times$ strategy finances one dollar for each dollar of investor capital. A $3\times$ strategy finances two. In the 1980s, the average annualized daily risk-free rate in the sample was $8.49\%$. The corresponding average financing charge was $8.49\%$ per year at $2\times$ and $16.98\%$ at $3\times$, before fund expenses.

Table~\ref{tab:longrunsimulation} reports the realized results through December 31, 2024. With financing but no fund fee, each levered strategy finishes ahead of the unlevered market. A common $1.47\%$ fee reduces terminal wealth by roughly half. It leaves the $1.25\times$ strategy slightly behind the unlevered market, while the higher-leverage strategies remain ahead in this sample. One sample window cannot isolate a single leverage effect because realized compounding, borrowing rates, fund expenses, and leverage all matter.

\begin{table}[htbp]
\centering
\caption{Daily-reset U.S. equity-market leverage since August 31, 1976}
\label{tab:longrunsimulation}
\begin{tabular}{rccccc}
\toprule
& \multicolumn{2}{c}{Financing only} &
\multicolumn{2}{c}{Financing plus 1.47\% fee} & \\
\cmidrule(lr){2-3}\cmidrule(lr){4-5}
$L$ & Ending wealth & CAGR & Ending wealth & CAGR & Max. drawdown \\
\midrule
$1.00$ & $235.31$ & $11.95\%$ & $235.31$ & $11.95\%$ & $-54.7\%$ \\
$1.25$ & $446.00$ & $13.44\%$ & $219.13$ & $11.79\%$ & $-65.3\%$ \\
$1.50$ & $771.90$ & $14.74\%$ & $379.26$ & $13.06\%$ & $-74.8\%$ \\
$2.00$ & $1{,}755.04$ & $16.70\%$ & $862.28$ & $15.00\%$ & $-87.8\%$ \\
$3.00$ & $2{,}921.43$ & $17.94\%$ & $1{,}435.07$ & $16.22\%$ & $-97.9\%$ \\
\bottomrule
\end{tabular}

\vspace{0.5em}
\begin{minipage}{0.92\textwidth}
\footnotesize
\emph{Notes:} The sample ends December 31, 2024. The market return is the dividend-inclusive Fama--French market return, calculated as $\mathrm{MKT-RF}+\mathrm{RF}$. Financing uses the daily Fama--French risk-free rate. The fee is subtracted only when $L>1$ and is a sensitivity using ULPIX's current expense ratio, not a historical fund backtest. The calculation excludes trading costs, taxes, spreads, and leverage constraints. Maximum drawdown is reported for the fee sensitivity.
\end{minipage}
\end{table}

\subsection{Product selection and clientele}
\label{app:clientele}
\par

In \citet{MurraySammon2026}, we find that long broad-equity-index LETFs generated more than \$100 billion in investor gains in their sample, including more than \$40 billion relative to counterfactual investments that placed the same flows at the same times in the underlying assets. Those products combined strong underlying returns with moderate volatility and relatively low financing costs.

Single-stock LETFs combine a different set of product characteristics. These funds are often launched on volatile recent winners, and financing spreads are higher on average. Volatility and costs raise the breakeven return, while product selection determines where investor dollars receive leverage. The observed outcomes therefore depend on which products exist and attract assets, not on the daily-reset formula alone.

The investors using a product can also change over time. A severe loss can change demand for the product and the composition of the investors who remain. More generally, a post-crisis sample can disproportionately represent investors whose strategies and preferences survived the crisis, while a product with a long upward history may attract and retain investors who extrapolate recent returns. A future bear market can therefore change both realized performance and the clientele whose behavior is observed afterward. This selection is part of interpreting flow-weighted outcomes. It is not a mechanical property of the LETF payoff.

\subsection{A moderate-leverage, low-cost design}
\label{app:moderate_leverage}

There is an economic reason to consider equity exposure above $1\times$ without jumping to $2\times$ or $3\times$ leverage. Early in working life, human capital can be large relative to financial wealth. Investing only current financial wealth in equities may therefore leave lifetime resources with relatively little equity exposure. Standard life-cycle models can support levered exposure for some young investors \citep{samuelson1969lifetime,merton1969lifetime}, and \citet{ayres2014practical} discuss its practical application. This argument does not imply that high leverage, expensive leverage, or leverage applied to a volatile individual asset is desirable.

For a long-run investor, this motivation points toward a particular product design: a broad equity index, moderate leverage, a very low expense ratio, and inexpensive financing. To make those terms concrete, Table~\ref{tab:moderateleverage} simulates daily-reset $1.25\times$ and $1.5\times$ exposure to the broad U.S. equity market. The strategies pay the daily risk-free rate on their incremental exposure and a five-basis-point annual fee. Five basis points is slightly above the $0.03\%$ expense ratios currently reported for \href{https://investor.vanguard.com/investment-products/etfs/profile/vti}{VTI} and \href{https://investor.vanguard.com/investment-products/etfs/profile/voo}{VOO}. The levered return follows Equation~\eqref{eq:financingfees} with $f=0.0005$. The $1\times$ comparison is the dividend-inclusive Fama--French market return and pays no product fee.

\begin{table}[htbp]
\centering
\caption{Effects of moderate leverage across sample periods}
\label{tab:moderateleverage}
\scriptsize
\setlength{\tabcolsep}{3.5pt}
\begin{tabular}{lrrrrrr}
\toprule
& \multicolumn{2}{c}{$1\times$} &
\multicolumn{2}{c}{$1.25\times$} &
\multicolumn{2}{c}{$1.5\times$} \\
\cmidrule(lr){2-3}\cmidrule(lr){4-5}\cmidrule(lr){6-7}
Sample & CAGR & Max. DD & CAGR & Max. DD & CAGR & Max. DD \\
\midrule
Aug. 1976--Dec. 2024 & $11.95\%$ & $-54.7\%$ & $13.39\%$ & $-64.2\%$ & $14.68\%$ & $-72.0\%$ \\
Sep. 1998--Dec. 2024 & $9.53\%$ & $-54.7\%$ & $10.79\%$ & $-64.2\%$ & $11.85\%$ & $-72.0\%$ \\
Jun. 2006--Dec. 2024 & $10.97\%$ & $-54.7\%$ & $12.73\%$ & $-64.2\%$ & $14.29\%$ & $-72.0\%$ \\
Feb. 2010--Dec. 2024 & $14.34\%$ & $-34.3\%$ & $17.27\%$ & $-41.5\%$ & $20.08\%$ & $-48.1\%$ \\
Jan. 2000--Dec. 2009 & $-0.38\%$ & $-54.7\%$ & $-1.97\%$ & $-64.2\%$ & $-3.78\%$ & $-72.0\%$ \\
\bottomrule
\end{tabular}

\vspace{0.5em}
\begin{minipage}{0.94\textwidth}
\footnotesize
\emph{Notes:} The levered strategies reset daily, pay a $0.05\%$ annual fee, and finance incremental exposure at the daily Fama--French risk-free rate. The $1\times$ benchmark is the dividend-inclusive Fama--French market return, $\mathrm{MKT-RF}+\mathrm{RF}$, and pays no fee. The calculations exclude funding spreads, trading costs, taxes, and leverage constraints. They describe realized historical paths, not investable fund returns or expected future performance.
\end{minipage}
\end{table}

Moderate leverage raises realized CAGR in each of the four long samples ending in 2024. It does not do so in every market environment. During 2000--2009, the unlevered market lost $0.38\%$ per year. The losses were $1.97\%$ at $1.25\times$ and $3.78\%$ at $1.5\times$. Drawdowns also become materially larger as leverage rises. The same additional exposure that helps in a long rising market compounds losses when weak returns and high volatility occur together.

These historical paths do not identify an optimal leverage ratio. They show the conditions under which moderate leverage is more plausible as a long-run product. Lower leverage reduces both the amount borrowed and the effect of squared returns in Equation~\eqref{eq:costloggap}. A five-basis-point fee avoids the large recurring expense in the ULPIX and UOPIX example, while financing at the risk-free rate is deliberately favorable. An actual product would also face funding spreads, tracking error, trading costs, and taxes. Even with a broad index and low costs, suitability depends on the investor's horizon, other assets, human capital, liquidity needs, and ability to bear the larger drawdowns.

\renewcommand{\bibfont}{\footnotesize}

\end{document}